\documentclass[12pt,preprint]{aastex}

\shorttitle{ Circular orbits and accretion process in
Horndeski/Galileon gravity}\shortauthors{Salahshoor, Nozari}

\begin{document}

\title{Circular orbits and accretion process in a class of Horndeski/Galileon black holes}
\author{K. Salahshoor$^{a,}$\footnote{k.salahshoor@stu.umz.ac.ir}\,\, and \,\, K. Nozari$^{a,b,}$\footnote{knozari@umz.ac.ir(Corresponding Author)}}

\affil{$^{a}$Department of Physics, Faculty of Basic Sciences,\\ University of Mazandaran, P. O. Box 47416-95447, Babolsar,
IRAN}

\affil{$^{b}$Research Institute for Astronomy and Astrophysics of Maragha
(RIAAM),\\ P. O. Box 55134-441, Maragha, Iran}

\begin{abstract}
In this paper the geodesics motion and accretion process around a
subclass of Horndeski/Galileon black holes are investigated. Firstly, we present
spherically symmetric geometries in a Horndeski/Galileon black hole spacetime by
considering an isothermal fluid around the black hole. Then we focus
on three main issues: in the first step circular orbits of test particles and
their stability in equatorial plane are examined in details.
Then, by treating perturbations via restoring forces, oscillations of particles around the central object are studied.
Finally, the accretion process, the critical speed of the flow and
accretion rate are investigated in this setup properly.
\end{abstract}


\section{Introduction}
Recent observations indicate that General Relativity might indeed be
modified at large distances. Scalar-tensor theories are a prototype
alternative and also they are most probably the simplest, consistent
and nontrivial modification of the General Relativity (Babichev \emph{et
al.} 2016). Gregory Horndeski (Horndeski 1974) proposed the most
general action of the scalar-tensor gravity. The same results were
formulated in terms of Galileons interactions (Deffayet
2011, Kobayashi 2011, Deffayet 2013). Therefore,
Horndeski/Galileons is the most general class of scalar-tensor field
models with second-order field equations and it may be considered as
a proper generalization of General Relativity in high energy regime
(Latosh 2016). We consider the following action which is a subclass of general Horndeski/Galileons class

\begin{equation}\label{a1}
S=\int d^{4}x\sqrt{-g}[\zeta R-\eta(\partial\phi)^{2}+\beta
G^{\mu\nu}\partial_{\mu}\phi \partial_{\nu}\phi-2\Lambda].
\end{equation}

Here $R$ is the Einstein-Hilbert term, $G^{\mu\nu}$ is the Einstein tensor, $\phi$ is the scalar
field,  $\Lambda$ is a cosmological constant term, $\zeta>0$,
$\eta$ and $\beta$ are model parameters. On astrophysical scales, which covers also static
and spherically symmetric solutions, this scalar-tensor
theory may play a crucial role. On the other hand, any modification of General Relativity must be
consistent with astrophysical observations. So, it is important to see how astrophysical processes such as
accretion onto black holes work in this scalar-tensor framework and can be
used also as a probe to see viability of these theories from experimental viewpoint.
The issue of black hole accretion disk is studied in some subclasses of
scalar-tensor theories. However, there is a gap in literature since black hole accretion disk
has not been studied in the mentioned scalar-tensor theory as a subclass of the general Horndeski/Galileon scenario.
This is the motivation of the present study and we are going to fill this gap in this paper.

Accretion disks are constructed by rotating gaseous materials that move
in bounded orbits because of the gravitational force of central
mass, such as Young Stellar Objects (YSO), main-sequence stars
(MSs), neutron stars (NSs), and supermassive black holes in Active
Galactic Nuclei (AGN). In such systems, particles orbits are
stable, but when the orbits of these materials become unstable,
following it, accretion will be happened. Accretion is the process
by which a massive central object such as a black hole captures
particles from a fluid in its vicinity. The particles which
accelerate from rest must be passed through a critical point, the
point where the velocity of the gas matches its local sound speed.
Then the gas falls onto the central mass at supersonic velocities.
This process leads to increase in mass of the black hole (Martnez 2014).
In addition, extra energy would be released in this process
where this energy can be source of some astrophysical phenomena, such
as the production of powerful jets, high-energy radiation, and
quasars (Kato \emph{et al.} 2008). Therefore, the study of the geodesic
structure of particles in the vicinity of black holes and specially
investigation about some characteristic radii such as marginally
bound orbits $(r_{mb})$ and innermost stable circular orbits $(r_{isco})$
are interesting issues for a careful study of the subject matter. These radii are very
important in the study of black hole accretion disks. For example,
in thin accretion disks, the inner edge of the disk coincides with the
innermost stable circular orbit (ISCO) and the efficiency of the
energy released, which describes the significance of converting
rest-mass energy into radiative energy (Xie \emph{et al.} 2012), can be
determined from this radii.

The location of unstable or stable circular orbits is consistent
with the maximum or minimum of the effective potential
respectively. In Newtonian theory, for any value of the angular
momentum, the effective potential has a minimum and then
stable circular orbit is free to have arbitrary radius, that is,
there is no minimum radius of stable circular orbit, (ISCO)
(Kaplane 1949). But this situation is different when the effective
potential has a complicated form depending on the particle angular
momentum and other parameters or when one incorporates general
relativistic effects. For example, in General Relativity and for
particles moving around the Schwarzschild black hole, for any value
of the angular momentum, the effective potential has two extrema
(minimum or maximum). But, only for a specific value of the angular
momentum the two points coincide. This point introduces ISCO where
is located at $r=3r_{g}$ (Landau 1993, Kaplane 1949) where $r_{g}$
is the Schwarzschild radius. In different metrics, the properties of
spacetime affects the locations of these radii and some parameters
such as: specific energy, angular momentum and angular velocity are
important in the position of these points. A lot of research programs are
devoted to study these radii and their physical significance.
Ruffini \emph{et al.} (1971) and Bardeen \emph{et al.} (1972) studied the
properties of innermost stable circular orbits around the Kerr black
hole.  Even Hobson \emph{et al.} (2006) described these features in
details in their textbook on General Relativity. The radiation efficiency of accretion
disks, $\eta$, for Schwarzschild and Kerr black holes was obtained
by Novikov and Thorne (1973) which its value lies in the range of
$0.057$ $-$ $0.43$ depending on the black hole spin. The Kerr-like
metric was constructed by Johannsen and Psaltis (2011) and then
Johannsen (2013) has studied the accretion disks around such black
holes. The study of the geodesic motion and the circular orbits of charged
particles around weakly magnetized rotating black holes are carried out
by Tursunov \emph{et al.} (2016).

In an accretion disk particles move in stable orbits but when a
perturbations, as a result of restoring forces, act on the particles,
oscillations around the circular orbit can take place in vertical
and radial directions with epicyclic frequencies.
Happening the oscillations (in response to perturbations) in the inner region of an accretion disk is
another important characteristic of these regions. Oscillations can be
source of strong and chaotic time variations in spectrum of such
systems. Therefore, study about orbital and epicyclic frequencies (radial and vertical) play
an important role in the physics of relativistic accretion disks around the black holes.
Isper (1994, 1996), Wagoner (1999), Kato (2001) and Ortega-Rodriges \emph{et al.} (2006) have studied in
this field. Resonance between such a frequency modes which proposed
by Kluzniak and Abromowicz (2000) can be a physical mechanism for
existing Quasi-periodic oscillations (QPOs). QPOs in the X-ray
fluxes of some astrophysical objects such as a neutron star and black
hole sources have been reviewed by many researches including van der
Klis (2000) and McClintock \emph{et al.} (2003). Johannsen (2013) has examined
the radial and vertical epicyclic frequencies in the Kerr-like
metric.

With these preliminaries, in this paper we study nonrotating black hole solutions with accretion disk in a subclass
of Horndeski/Galileons spacetime general class. For
simplicity we restrict our study to equatorial plane in a polar
coordinates system. Firstly, the singularity and event horizon in
this spacetime geometry are presented. Then, in order to investigate
the circular orbits, effective potential is obtained in this setup.
We study the locations of several characteristic radii, such as:
marginally stable circular orbits $r_{isco}$, marginally bounded
circular orbits $r_{mb}$ and photon orbits $r_{ph}$ in equatorial
plane. Also, the ISCO binding energy, the maximum radiation
efficiency, the emission and temperature in equatorial epicyclic
frequencies are computed. Finally, some dynamical parameters and
critical accretion of isothermal fluid are investigated in details.

This paper is organized as follows: In section 2 we introduce
Horndeski/Galileon spacetime. The general formalism of a test
particle's motion is discussed in section 3 where circular motion,
stable circular orbits and oscillations are examined in subsections
3.1, 3.2 and 3.3. In section 4 and it's subsections, the general
form of some dynamical parameters such as critical speed of the
flow, accretion rate and the time of accretion for an isothermal
fluid are obtained. In section 5 we have explained physically all of
these results for a subclass of general solution of the
Horndeski/Galileon black hole. Finally, section 6 is devoted to
summary and discussion.

\section{Horndeski/Galileon Spacetime }

We study static and spherically symmetric
limit of black hole solutions in a subclass of general Horndeski/Galileon theories (see Maselli (2015)
for the case of slowly rotating black holes in Horndeski theory and also Babichev et. al. (2016) for black hole and star solutions for Horndeski theory). The general form of the line-element for such
systems with the metric signature $(+, -, -, -)$ is described by
\begin{equation}\label{a1}
ds^{2}=h(r)dt^{2}-\frac{1}{f(r)}dr^{2}-r^{2}(d\theta^{2}+\sin^{2}\theta
d\varphi^{2}).
\end{equation}
The metric functions, $f(r)$ and $h(r)$, depend only on the radial
coordinate $r$ and are given as follows (Tretyakova 2016)
\begin{eqnarray}\label{a2}
f(r)=\frac{(\beta+\eta r^2)h(r)}{\beta (r h(r))'},
\end{eqnarray}
\begin{equation}\label{a1}
h(r)=-\frac{\mu}{r}+\frac{1}{r}\int \frac{k(r)}{(\beta+\eta r^2)}dr,
\end{equation}
\begin{equation}
\phi(r)=qt+\psi(r)\,.
\end{equation}
In these relations, $\mu$ plays the role of the mass term and $k$ is
obtained from the following constraint equation
\begin{equation}\label{a1}
q^{2}\beta(\beta+\eta r^2)^{2}-[2\zeta \beta+(2\zeta
\eta-\lambda)r^{2}]k+C_{0}k^{\frac{3}{2}}=0\,,
\end{equation}
where $C_{0}$ is a constant of integration. It is important to note that the static metric (2)
has the time rescaling symmetry. So, if $h(r)$ is a solution then $c h(r)$ should be also
a solution where $c$ is a constant. This means that $c$ should be set for $t$ to describe
the proper time in the relevant region, namely $h \rightarrow 1$.
We note also that equation (5) is not a solution, but rather an ansatz on the
scalar field. This type of scalar configuration was firstly considered by
Babichev and Charmousis (2014) in which the solution for $\psi(r)$ is also presented.
About the stability of solutions, Ogawa \emph{et al.} (2016) and  Takahashi and Suyama (2017)
showed that solutions with nonzero $q$ are generically plagued by ghost or gradient instability.
However, for solutions with $q=0$, there exist some stable solutions (see Kobayashi \emph{et al.} (2012),
Takahashi and Suyama (2017), and Tretyakova and Takahashi (2017)). We note that recently it has been pointed out by (Babichev et. al. (2017) and (2018)) that the conclusion of Ogawa et al. (2016) and Takahashi and Suyama (2017) is incorrect. Now from the recent relations, various solutions for different values of $C_{0}$ and $q$ can be
obtained. We are going to discuss a common expression in this
paper.

Now we study the properties of a subclass of black hole solutions in Horndeski/Galileon gravity as has been introduced above. The mentioned
Horndeski spacetime contains a singularity at the location where
the following condition holds
\begin{eqnarray}\label{a2}
f(r)=0\,.
\end{eqnarray}
The event horizon is a null surface. A surface that is defined as $f(x^{\mu})=0$ will be null if
\begin{equation}\label{a1}
g^{\mu\nu}n_{\mu} n_{\nu}=0,
\end{equation}
where $n_{\mu}$ is the normal 4-vector to the surface and it is defined as
$n_{\mu}=\nabla_{\mu}f$. Since we are interested in to study the problem in the equatorial plane,
then the relation (8) can be written as $g^{rr}(\partial_{r}f)^{2}=0$. Therefore, in the
radial distance that $g^{rr}=0$ or equivalently $f(r)=0$, we would have an
event horizon. The location of the event horizon in the geometry is a radial distance from the center of the core
where the metric is singular, except the intrinsic singularity which
cannot be removed via coordinate transformation.


\section{Test Particle's Motion: General Formalism}
The motion of a test particle is governed by the geodesic structure
of the underlying spacetime manifold. In this section, we study
general form of timelike geodesics around a subclass
of Horndeski/Galileon black hole. Spacetime around this
black hole is static and symmetric with two Killing vectors $\xi_{t}
=\partial_{t} $ and $\xi_{\varphi}=\partial_{\varphi} $ which imply
two constants of motion $E$ and $L$ (conserved energy and angular
momentum per unit mass) along the trajectory as follows
\begin{eqnarray}\label{a2}
&E=-g_{\mu\nu}\xi^{\mu}_{t} u^{\nu}\equiv -u_{t}\,,\nonumber\\
&L=g_{\mu\nu}\xi^{\mu}_{\varphi} u^{\nu}\equiv u_{\varphi}\,,
\end{eqnarray}
where $u^{\mu}=(u^{t}, u^{r}, u^{\theta}, u^{\varphi})$ is the
four-velocity of the test particle. Using the normalization
condition for four-velocity, that is $u^{\mu}u_{\mu}=1$,  we have
\begin{equation}\label{a3}
[g_{rr}(u^{r})^{2}+g_{\theta\theta}(u^{\theta})^{2}]=[1-g^{tt}(u_{t})^{2}-g^{\varphi\varphi}(u_{\varphi})^{2}]\,.
\end{equation}
From equations (9) and  (10) and in equatorial plane with
$\theta=\frac{\pi}{2}$, four-velocity will be given by the following components
\begin{eqnarray}\label{a6}
&u^{t}=-\frac{E}{h(r)}\nonumber\\
&u^{\theta}=0\nonumber\\
&u^{\varphi}=-\frac{L}{r^{2}}\nonumber\\
&u^{r}=[-f(r)(1-\frac{E^{2}}{h(r)}+\frac{L^{2}}{r^{2}})]^{\frac{1}{2}}\,.
\end{eqnarray}
Also the following equation can be derived easily
\begin{equation}\label{a4}
\frac{h(r)}{f(r)}(u^{r})^{2}+V_{eff}=E^{2}\,.
\end{equation}
In this equation $V_{eff}$ is the effective potential for the test
particle motion that is given by
\begin{equation}\label{a5}
V_{eff}=h(r)\Big[1+\frac{L^{2}}{r^{2}}\Big]\,.
\end{equation}
It is clear that effective potential depends on the particle's
specific angular momentum radial distribution and the spacetime
parameter via $h(r)$. The study of effective potential is
very useful in geodesic motion. For example, the local
exterma of the effective potential determine the location of the
circular orbits.

\subsection{Circular Motion }
For circular motion in the equatorial plane, radial component $r$
must be constant and so $u^{r}=\dot{u}^{r}=0$ must be satisfied.
 Therefore, from equation (12) we would have $V_{eff}=E^{2}$ and
$\frac{d}{dr}V_{eff}=0$. From these relations the specific energy
$E$, the specific angular momentum $L$, the angular velocity
$\Omega_{\varphi}$ and angular momentum $l$ can be obtained by the
following relations respectively
\begin{equation}\label{a7}
E^{2}=\frac{2h^{2}(r)}{2h(r)-r h'(r)}\,,
\end{equation}
\begin{equation}\label{a8}
L^{2}=\frac{r^{3}h'(r)}{2h(r)-r h'(r)}\,,
\end{equation}
\begin{equation}\label{a9}
\Omega_{\varphi}=\frac{d\varphi}{dt}\equiv
\frac{u^{\varphi}}{u^{t}}\Rightarrow
\Omega_{\varphi}^{2}=\frac{1}{2r}h'(r)\,,
\end{equation}
\begin{equation}\label{a10}
l^{2}=\frac{L^{2}}{E^{2}}=\frac{r^{3}}{2h^{2}(r)}h'(r)\,.
\end{equation}
In order the energy and angular momentum to be real, the following
condition must be satisfied
\begin{eqnarray}\label{a11}
2h(r)-r h'(r)>0\,.
\end{eqnarray}
By solving this inequality, the limited area of circular orbit can
be obtained. Therefore,  this is the condition for existence of the
circular orbits. For bound orbit the relation $E^{2}<1$ must be hold
and in marginally bound orbits we have $E^{2}=1$. Then from equation
(14) we fined
\begin{eqnarray}\label{a12}
rh'(r)+2h(r)[h(r)-1]=0
\end{eqnarray}
By solving this equation, marginally bound orbits can be obtained easily.
From equations (14) and (15) it is seen that the energy and angular momentum diverge at the
radius where the following relation holds
\begin{eqnarray}\label{a11}
2h(r)-r h'(r)=0\,.
\end{eqnarray}
Photon sphere can be obtained by solving this relation. In a
photon sphere, photon moves on circular orbits. This region plays a
crucial role in the study of gravitational lensing, since lensing
effect cannot be observed below this region.


\subsection{Stable Circular Orbits and Radiant Energy Flux}

The local minima  of the effective potential correspond to the
stable circular orbits. Thus a stable circular orbit exists if
$\frac{d^{2}}{dr^{2}}V_{eff}>0$ and in addition to this condition,
in marginally stable circular orbits, $r_{isco}$, the condition
$\frac{d^{2}}{dr^{2}}V_{eff}=0$ must be satisfied. From equation (13) we
have
\begin{eqnarray}\label{a11}
\frac{d^{2}}{dr^{2}}V_{eff}=h''(r)(1+\frac{L^{2}}{r^{2}})-4h'(r)\frac{L^{2}}{r^{3}}+6h(r)\frac{L^{2}}{r^{4}}\,.
\end{eqnarray}
Accretion process is possible in  $r<r_{isco}$. When falling
particles from rest at infinity accrete onto the central mass, the
released gravitational energy of falling particles can convert into
radiation where this energy is the source of the most energetic
phenomena in astrophysics. The flux of the radiant energy over the
disk can be expressed in terms of the specific angular momentum $L$,
the specific energy $E$ and the angular velocity $\Omega_{\varphi}$
by the following relation (see for instance Kato et al., (2008))
\begin{eqnarray}\label{a13}
K=-\frac{\dot{M}\Omega_{\varphi,r}}{4\pi \sqrt{-g}(E-L
\Omega_{\varphi})^{2}}\int_{r_{ms}}^{r}(E-L
\Omega_{\varphi})L_{,r}dr\,,
\end{eqnarray}
where $\dot{M}$ is the accretion rate, $\Omega_{\varphi,r}\equiv
\frac{d\Omega_{\varphi}}{dr}$ and the parameter $g$ is determinant
of $g_{\mu\nu}$ given by
\begin{eqnarray}\label{a13}
g=det(g_{\mu\nu})=-\frac{h(r)}{f(r)}r^{4}\sin^{2}\theta\,.
\end{eqnarray}
We set $\sin\theta=1$, since we restrict our studies in  equatorial
plan. From relations (14)-(16) we would have
\begin{eqnarray}\label{a13}
K(r)=-\frac{\dot{M}}{4\pi r^{4}}\sqrt{\frac{r f(r)}{2
h(r)h'(r)}}\Big(\frac{[2 h(r)-r h'(r)][r h''(r)-h'(r)]}{[2 h(r)+r
h'(r)]^{2}}\Big)\int_{r_{ms}}^{r}\mathcal{F}(r)dr\,,
\end{eqnarray}
where by definition
\begin{eqnarray}\label{a13}
 \mathcal{F}(r)\equiv\sqrt{\frac{r}{2h'(r)}}\frac{[2 h(r)+r
h'(r)][- h''(r)r h(r)+2r h'^{2}(r)-3h'(r) h(r)]}{[2 h(r)-r
h'(r)]^{2}}\,.
\end{eqnarray}

The steady-state accretion disk model is supposed to be in
thermodynamical equilibrium. Then the radiation emitted from the
surface of the disk can be as a black body radiation. So, the
relation $K(r)=\sigma T^{4}(r)$ can be hold between energy flux
emitted at the surface of the disk and effective temperature of the
disk ($\sigma$ is the Stefan-Boltzman constant). Using this
relation, temperature distribution on the disk by assuming thermal
black body radiation can be obtained easily and then we can compute
the luminosity $L(\nu)$ of the disk. The observed
luminosity at the distance $d$ to the source with the disk inclination
angle $\gamma $ has the following form (Torres  2002)
\begin{eqnarray}\label{a58}
L(\nu)=4 \pi d^{2} I(\nu)=\frac{8}{\pi}(\cos \gamma)\int
_{r_{i}}^{r_{f}}\int_{0}^{2\pi} \frac{\nu_{e}^{3}\,r\, d\varphi\,\,
dr}{\exp(\frac{\nu_{e}}{T})-1}\,,
\end{eqnarray}
where $I(\nu)$ is the thermal energy flux. In this relation $r_{i}$
indicates the position of the inner edge and we take $r_{i}=r_{ms}$.
Also $r_{f}$ indicates the outer edge of the disk. Since for
any kind of general relativistic compact object the flux over the
disk surface could be vanishing at $r\rightarrow \infty$, we take
$r_{f}\rightarrow \infty$. The emitted frequency is given by
$\nu_{e}=\nu(1+z)$ where the redshift factor $z$, by neglecting the light
bending, can be written as follows
\begin{eqnarray}\label{a58}
z=\frac{1+\Omega_{\varphi} r \sin \varphi\, \sin
\gamma}{\sqrt{-g_{tt}-\Omega_{\varphi}^{2}g_{\varphi\varphi}}}-1\,.
\end{eqnarray}

The efficiency of the accreting flow is another important characteristic of the mass accretion
process. The maximum efficiency of transforming gravitational energy
into radiative flux of such particles between innermost circular
orbit and infinity, $\eta^{*}$, is defined as the ratio of the specific
binding energy of the innermost circular orbit to the specific rest
mass energy which is given by the following relation
\begin{eqnarray}\label{a58}
\eta^{*}=1-E_{isco},
\end{eqnarray}
where $E_{isco}$ is the specific energy of a particle rotating in an
innermost stable circular orbit. This relation is valid for the case
where all the emitted photons can escape to infinity.

Now we focus on perturbations. If a perturbation acts on  the fluid element, the motion of a test
particle will be nearly circular orbit in the equatorial plane and
the particle will oscillate around the circular orbit with three
components of motion, the issue which is discussed in the next section.

\subsection{Oscillations }
In an accretion disk, various types of oscillatory motions as a
result of restoring forces are expected. Restoring forces act on
perturbations in the accretion disks resulting Horizontal and
Vertical oscillations. Some of these restoring forces in accretion
disks are resulting from rotation of the disk and from a vertical
gravitational filed. When a fluid element is displaced in the radial direction, it will
return to its equilibrium position due to a restoring force
resulting from rotation of the fluid. In accretion disks, because
of existence of central object, centrifugal force is balanced by the
gravitational force. When the former dominates over the latter or
the reverse happens, the element of flow will be pushed inward or
outward to return to the original radius with epicyclic frequency
$\Omega_{r}$. On the other hand, when a fluid element is perturbed
in the vertical direction, the vertical component of the
gravitational field returns the perturbed element toward equilibrium
position, that is, the equatorial plane. As a result of this
restoring force, the element of the fluid makes harmonic oscillation
around the equatorial plane with vertical epicyclic oscillations
$\Omega_{\theta}$ (Kato \emph{at al.} 2008).

In a general relativistic discussion about the motion of the fluid
in an accretion disk, three frequencies around the central object
are important. Circular motion at the orbital frequency
$\Omega_{\varphi}$, harmonic radial motion at the radial frequency
$\Omega_{r}$ and the harmonic vertical motion at the vertical
frequency $\Omega_{\theta}$. As we have stated, resonance between
such frequencies can be source of quasi-periodic oscillations which
leads to chaotic and quasi-periodic variability in X-ray fluxes
from many galactic black holes. Study in this field is
important in some sense. For this purpose, radial and vertical
motions around a circular equatorial plane are discussed in this
section.

Radial and vertical motions can be explained by $\frac{1}{2}(\frac{dr}{dt})^{2}=V^{(r)}_{eff}$
and $\frac{1}{2}(\frac{d\theta}{dt})^{2}=V^{(\theta)}_{eff}$ where
from equation (10), to describe radial motion $u^{\theta}=0$, and
also for describing the vertical motion we have $u^{r}=0$. By
setting $u^{r}=\frac{dr}{d\tau}=\frac{dr}{dt}u^{t}$ and
$u^{\theta}=\frac{d\theta}{d\tau}=\frac{d\theta}{dt}u^{t}$,  we find
\begin{eqnarray}\label{a13}
&\frac{1}{2}(\frac{dr}{dt})^{2}=-\frac{1}{2}\frac{f(r)h^{2}(r)}{E^{2}}[1-\frac{E^{2}}{h(r)}+\frac{L^{2}}{r^{2}sin^{2}\theta}]=V^{(r)}_{eff}\,,\nonumber\\
&\frac{1}{2}(\frac{d\theta}{dt})^{2}=-\frac{1}{2}\frac{h^{2}(r)}{r^{2}E^{2}}[1-\frac{E^{2}}{h(r)}+\frac{L^{2}}{r^{2}sin^{2}\theta}]=V^{(\theta)}_{eff}\,.
\end{eqnarray}
In order to investigate the radial and vertical epicyclic
frequencies, small perturbations  $\delta r$ and $\delta \theta$
around the circular orbit in equatorial plane are considered.
By taking the time-derivative of the first equation in (29), equation
describing the radial oscillations can be obtained as follows
\begin{eqnarray}\label{a14}
\frac{d^{2}r}{dt^{2}}=\frac{dV_{eff}^{(r)}}{dr}.
\end{eqnarray}
For a particle which is perturbed from its original radius at
$r=r_{0}$ by a deviation  $\delta r=r-r_{0}$, the perturbed equation of
motion is given by
\begin{eqnarray}\label{a14}
\frac{d^{2}(\delta
r)}{dt^{2}}=\frac{d^{2}V_{eff}^{(r)}}{dr^{2}}(\delta r)\Rightarrow
(\delta {\ddot r})+\Omega^{2}_{r}(\delta r)=0\,,
\end{eqnarray}
where a dote denotes differential with respect to time coordinate
$t$ and $\Omega_{r}^{2}\equiv-\frac{d^{2}V_{eff}^{(r)}}{dr^{2}}$. By the
same procedure, for a perturbation in the vertical direction by a
deviation given as $\delta \theta=\theta-\theta_{0}$ we find
\begin{eqnarray}\label{a14}
\frac{d^{2}(\delta
\theta)}{dt^{2}}=\frac{d^{2}V_{eff}^{(\theta)}}{dr^{2}}\delta
\theta\Rightarrow (\delta \ddot{\theta})+\Omega^{2}_{\theta}(\delta
\theta)=0\,,
\end{eqnarray}
where $\Omega^{2}_{\theta}=-\frac{d^{2}}{d\theta^{2}}V^{(\theta)}_{eff}$.
Then from equations (29) in equatorial plane we would have respectively
\begin{eqnarray}\label{a15}
&\Omega^{2}_{r}=\frac{1}{2r^{4}E^{2}}\Big\{\Big[(r^{2}+L^{2})r^{2}h^{2}(r)-h(r)r^{4}E^{2}\Big]f''(r)+\Big[(r^{2}+L^{2})2h(r)-r^{2}E^{2}\Big]r^{2}f(r)h''(r)\nonumber\\
&+2r^{2}f(r)h'^{2}(r)(r^{2}+L^{2})-2r\Big[[-(r^{2}+L^{2})2h(r)+r^{2}E^{2}]r f'(r)+4f(r)h(r)L^{2}\Big]h'(r)\nonumber\\
&-4h^{2}(r)L^{2}(-\frac{3}{2}f(r)+r f'(r)) \Big\}\,,
\end{eqnarray}
and
\begin{eqnarray}\label{a16}
\Omega^{2}_{\theta}=\frac{h^{2}(r)L^{2}}{r^{4}E^{2}}\,.
\end{eqnarray}
In these equations, a prime denotes differential with
respect to the radial coordinate, $r$. To proceed further, now we
present basic dynamical equations in this subclass of general Horndeski/Galileons black hole spacetime.


\section{Basic dynamical equations}
In this section we provide the basic dynamical equations for our forthcoming arguments (we refer to Babichev et al., (2005) and (2013) for more details).
Here we consider a perfect fluid which is specified by the
following energy-momentum tensor
\begin{eqnarray}\label{a18}
T^{\mu\nu}=(p+\rho)u^{\mu}u^{\nu}-p g^{\mu\nu},
\end{eqnarray}
where $p$ and $\rho$ are pressure and energy density of the fluid
respectively. In this relation, $u^{\mu}$ is the fluid elements
four-velocity. Because of background symmetries, in relation (35) all
of the components are functions of only the radial coordinate, $r$. Since we are assuming
the fluid is flowing radially in the equatorial plane
($\theta=\frac{\pi}{2}$),  the general form of the four-velocity
will be as follows
\begin{eqnarray}\label{a18}
u^{\mu}=\frac{dx^{\mu}}{d\tau}=(u^{t},u^{r},0,0)\,,
\end{eqnarray}
where $\tau$ is the proper time along the geodesic. From this
relation and under the normalization condition $u^{\mu}u_{\mu}=1$,
we obtain
\begin{eqnarray}\label{a18}
u^{t}=\sqrt{\frac{f(r)+(u^{r})^{2}}{h(r)f(r)}}\,,
\end{eqnarray}
where for forward flow in time, $u^{t}$ must be positive and for
accretion (inward flow), $u^{r}<0$. By deriving the energy-momentum and also particle-number
conservation equations, all of the required equations for studying
the accretion are obtained. Conservation of the energy- momentum
tensor is given by
\begin{eqnarray}\label{a18}
T^{\mu \nu}_{;\mu}=0\Rightarrow T^{\mu
\nu}_{;\mu}=\frac{1}{\sqrt{-g}}(\sqrt{-g}T^{\mu
\nu})_{,\mu}+\Gamma^{\nu}_{\alpha\mu}T^{\alpha \mu }=0\,,
\end{eqnarray}
where in this relation $(;)$ shows the covariant differentiation,
$\sqrt{-g}=r^{2}\sin \theta \sqrt{\frac{h(r)}{f(r)}}$ and $\Gamma$
is the second kind Christoffel symbol (affine connection) where
its non-zero components are as follows
\begin{eqnarray}\label{a18}
&\Gamma^{0}_{01}=\Gamma^{0}_{10}=\frac{1}{2}\frac{h'(r)}{h(r)}\nonumber\\
&\Gamma^{1}_{00}=\frac{1}{2}h'(r)f(r),\
\Gamma^{1}_{11}=-\frac{1}{2}\frac{f'(r)}{f(r)},\ \Gamma^{1}_{22}=-r
f(r), \    \Gamma^{1}_{33}=-r f(r)\sin \theta\nonumber\\
&\Gamma^{2}_{12}=\Gamma^{2}_{21}=\frac{1}{r}\nonumber\\
&\Gamma^{3}_{13}=\Gamma^{3}_{31}=\frac{1}{r}\,.
\end{eqnarray}
From these relations, equation (38) yields
\begin{eqnarray}\label{a18}
T^{10}_{,r}+\frac{1}{\sqrt{-g}}T^{10}(\sqrt{-g})_{,r}+2\Gamma^{0}_{01}T^{10}=0\,,
\end{eqnarray}
where after some manipulations we obtain
\begin{eqnarray}\label{a18}
(p+\rho)u^{r}r^{2}\sqrt{(u^{r})^{2}+f(r)}\frac{h(r)}{f(r)}=A_{0}\,,
\end{eqnarray}
with $A_{0}$ as an integration constant. Projecting the energy-momentum conservation law onto the
four-velocity via $u_{\mu}T^{\mu \nu}_{;\nu}=0$, yields
\begin{eqnarray}\label{a18}
(p+\rho)_{,\nu}
u_{\mu}u^{\mu}u^{\nu}+(p+\rho)u^{\mu}_{;\nu}u_{\mu}u^{\nu}+(p+\rho)u_{\mu}u^{\mu}u^{\nu}_{;\nu}+p_{,\nu}g^{\mu\nu}u_{\mu}+pu_{\mu}
g^{\mu\nu}_{;\nu}=0\,.
\end{eqnarray}
By considering the normalization conditions as  $u_{\mu}u^{\mu}=1$
and since $g^{\mu\nu}_{;\nu}=0$, this relation reduces to
\begin{eqnarray}\label{a18}
(p+\rho)u^{\nu}_{;\nu}+u^{\nu}\rho_{,\nu}=0\,.
\end{eqnarray}
Since $A^{b}_{;a}=\partial_{a}A^{b}+\Gamma^{b}_{ac}A^{c}$, we find
\begin{eqnarray}\label{a18}
u^{r}\rho_{,r}+(p+\rho)\Big[\Gamma^{0}_{0c}u^{c}+(u^{r}_{,r}+\Gamma^{1}_{1c}u^{c})+\Gamma^{2}_{2c}u^{c}+\Gamma^{3}_{3c}u^{c}\Big]=0\,.
\end{eqnarray}
By using the non-zero components of the connection, this relation after some
simplification yields
\begin{eqnarray}\label{a18}
\frac{\rho'}{(p+\rho)}+\frac{1}{2}\frac{h'(r)}{h(r)}-\frac{1}{2}\frac{f'(r)}{f(r)}+\frac{u'}{u}+\frac{2}{r}=0\,,
\end{eqnarray}
which after integration, we would have
\begin{eqnarray}\label{a18}
r^{2}u^{r}\sqrt{\frac{h(r)}{f(r)}} \exp(\int
\frac{d\rho}{p+\rho})=-A_{1}\,,
\end{eqnarray}
where $A_{1}$ is an integration constant. Since in the left hand side
$u^{r}<0$, the right hand side takes a minus sign too. So we find finally
\begin{eqnarray}\label{a18}
(p+\rho)\sqrt{h(r)\Big[\frac{(u^{r})^{2}}{f(r)}+1\Big]}\exp(-\int
\frac{d\rho}{p+\rho})=A_{2}\,,
\end{eqnarray}
where $A_{2}$ is an integration constant.
The equation of mass flux in this setup which is given by
\begin{eqnarray}\label{a18}
(\rho u^{\mu})_{;}\equiv \frac{1}{\sqrt{-g}}(\sqrt{-g} \rho
u^{\mu})_{,\mu}=0
\end{eqnarray}
 can be rewritten as
\begin{eqnarray}\label{a18}
\frac{1}{\sqrt{-g}}(\sqrt{-g} \rho
u^{\mu})_{,r}+\frac{1}{\sqrt{-g}}(\sqrt{-g} \rho
u^{\theta})_{,\theta}=0\,.
\end{eqnarray}
Since we are interested in to study just in equatorial plane, the
second term in equation (49) vanishes. Therefore, $\sqrt{-g} \rho
u^{\mu}$ would be as a constant, that is
\begin{eqnarray}\label{a18}
\rho u^{r}r^{2}\sqrt{\frac{h(r)}{f(r)}}=A_{3}\,,
\end{eqnarray}
where $A_{3}$ is an integration constant. We note that while we have
restricted our study to equatorial plane by symmetry considerations,
the general case is not so complicated in essence. Because of
symmetry all important characteristics of the model can be obtained
in $\theta=\frac{\pi}{2}$ plane as well. Now we are going to
determine dynamical parameter, critical accretion and accretion rate
in this setup.


\subsection{Dynamical parameters}

To proceed further, we assume isothermal fluids. These fluids flow
at a constant temperature. Therefore, $p\propto \rho$ and then the speed
of sound throughout the accretion process remains constant. For such
fluids, the equation of state is of the form $p=k \rho$ where $k$ is
the equation of state parameter. Then equations (46), (47) and (50)
yield
\begin{eqnarray}\label{a18}
\frac{p+\rho}{\rho}\sqrt{h(r)\Big[\frac{(u^{r})^{2}}{f(r)}+1\Big]}=A_{4}\,,
\end{eqnarray}
where $A_{4}$ is an integration constant. By substituting $p=k
\rho$, we can obtain $u$ as follows
\begin{eqnarray}\label{a18}
u=\Big(\frac{1}{k+1}\Big)\sqrt{f(r)\Big[\frac{A_{4}^{2}}{h(r)}-(k+1)^{2}\Big]}\,.
\end{eqnarray}
Then from eq. (50) the density of the fluid can be obtained as
\begin{eqnarray}\label{a18}
\rho=\frac{A_{3}}{r^{2}}\frac{(k+1)}{\sqrt{A_{4}^{2}-(k+1)^{2}h(r)}}\,.
\end{eqnarray}
Finally, from the relation $p=k \rho$, the pressure can be obtained
easily.


\subsection{Mass evolution}

In realistic astrophysical cases, the mass of the black hole is not
fixed in essence. By some processes such as accreting of mass from
accretion disk onto black hole and also Hawking radiation, its mass
will be changed gradually. The rate of change of mass can be
obtained by integrating the flux of the fluid over the surface of
the black hole, that is $\dot{M}\equiv\frac{dM}{dt}=-\int T^{r}_{t}
ds$ where a dot denotes the time derivative, $ds=\sqrt{-g}\,
d\theta\, d\varphi$ and $T^{r}_{t}=(p+\rho)u_{t}u^{r}$. By
substituting these relations, $\dot{M}$ can be obtained as follows
\begin{eqnarray}\label{a18}
\dot{M}=-4\pi r^{2}u(p+\rho)\frac{h(r)}{f(r)}\sqrt{u^{2}+f(r)}\equiv
-4\pi A_{0}\,,
\end{eqnarray}
where $A_{0}=-A_{1}A_{2}$, and $A_{2}=(p_{\infty}+\rho_{\infty})\sqrt{h(r_{\infty})}$. Therefore, we obtain
\begin{eqnarray}\label{a18}
\dot{M}=4\pi
A_{1}(p_{\infty}+\rho_{\infty})\sqrt{h(r_{\infty})}M^{2}.
\end{eqnarray}

Now, time evolution of mass of the black hole with initial mass $M_{i}$
can be obtained by integration of equation (55) that can be
rewritten as
\begin{eqnarray}\label{a18}
\frac{d M}{M^{2}}=\mathcal{F}t
\end{eqnarray}
where  $\mathcal{F}\equiv4\pi A_{1}(p+\rho)\sqrt{h(r_{\infty})}$. By
integrating from equation (56) we obtain
\begin{eqnarray}\label{a18}
M_{t}=\frac{M_{i}}{1-\mathcal{F}M_{i}t}\equiv
\frac{M_{i}}{1-\frac{t}{t_{cr}}}
\end{eqnarray}
where  $t_{cr}= \Big [4\pi A_{1}(p+\rho)\sqrt{h(r_{\infty})}M_{i} \Big ]^{-1}$ is the critical
accretion time. In the case $t=t_{cr}$ the denominator of equation (57) vanishes and the
black hole mass grows up to infinity in a finite time. After determining the
time evolution of the disk and black hole mass, now we study critical accretion in this setup.

\subsection{Critical Accretion}

Very far from the black hole, the flow is at rest but gravitational
field of black hole tends to accelerate it inwards. When flow moves
inwards, it must pass through critical point (sonic point) where in
this point $r=r_{c}$, the four-velocity of the fluid matches the
local speed of sound, $u=c_{s}$. In order to obtain sonic point, an expression for the radial
velocity gradient with no other derivatives is required. From
equations (50) and (51), the following two equations are obtained
\begin{eqnarray}\label{a18}
\frac{\rho'}{\rho}+\frac{u'}{u}+\frac{1}{2}\Big[\frac{h'(r)}{h(r)}-\frac{f'(r)}{f(r)}\Big]+\frac{2}{r}=0\,,
\end{eqnarray}
and
\begin{eqnarray}\label{a18}
\frac{\rho'}{\rho}\Big[\frac{d\ln(p+\rho)}{d\ln
\rho}-1\Big]+\frac{1}{2}\Big[\frac{h'(r)}{h(r)}-\frac{f'(r)}{f(r)}\Big]+\frac{u
u'}{u^{2}+f(r)}+\frac{1}{2}\frac{f'(r)}{u^{2}+f(r)}=0\,.
\end{eqnarray}
From these equations we obtain
\begin{eqnarray}\label{a18}
\frac{d\ln u}{d\ln r}=\frac{\mathcal{D}_{1}}{\mathcal{D}_{2}}\,,
\end{eqnarray}
where by definition
\begin{eqnarray}\label{a18}
\mathcal{D}_{1}=-\Big[\frac{r}{2}(V^{2}-1)\Big(\frac{h'(r)}{h(r)}-\frac{f'(r)}{f(r)}\Big)+2V^{2}-\frac{r
f'(r)}{2(u^{2}+f(r))}\Big]\,,
\end{eqnarray}
and
\begin{eqnarray}\label{a18}
\mathcal{D}_{2}=\Big[V^{2}-\frac{u^{2}}{u^{2}+f(r)}\Big]\,.
\end{eqnarray}
With these two equations, the following relation can be obtained
\begin{eqnarray}\label{a18}
V^{2}=\frac{d\ln (p+\rho) }{d\ln \rho}-1.
\end{eqnarray}
The condition for critical points is $\mathcal{D}_{1}=\mathcal{D}_{2}=0$. This condition yields
\begin{eqnarray}\label{a18}
V_{c}^{2}=\frac{r h'(r)}{4h(r)+rh'(r)}\,,
\end{eqnarray}
and
\begin{eqnarray}\label{a18}
u_{c}^{2}=\frac{1}{4}\frac{r f(r)h'(r)}{h(r)}\,,
\end{eqnarray}
where index $c$ refers to the critical point. Since the right hand side of the relation
(64) must be positive, then if dependence of $h(r)$ is known,  by
solving the following inequality the range of critical radius can
be obtained
\begin{eqnarray}\label{a18}
4h(r)+rh'(r)>0.
\end{eqnarray}
Finally the sound speed $c_{s}^{2}=\frac{dp}{d\rho}$ can be obtained
from (52) as
\begin{eqnarray}\label{a18}
c_{s}^{2}=A_{4}\sqrt{\frac{f(r)}{h(r)}\Big[u^{2}+f(r)\Big]}-1\,.
\end{eqnarray}

\section{ A Subclass of Horndeski/Galileon Black Hole Solutions }

As we have stated in section 2, a huge variety of solutions can be
obtained from equations (3)-(6). Here we present a solution of this subclass of the Horndeski/Galileon
setup that are characterized by the parameters $A, B, C$ and $\gamma$
(see for instance Tretyakova 2016). The solutions are as follows
\begin{equation}\label{a1}
h(r)=C-\frac{\mu}{r}+Ar^{2}+\Delta\,,
\end{equation}
and
\begin{equation}\label{a1}
f(r)=(1+\frac{\eta}{\beta}r^{2})\Big[\frac{\Delta+A
r^{2}+C-\frac{\mu}{r}}{3A
r^{2}+C+\frac{B}{1+r^{2}\gamma^{2}}}\Big]\,,
\end{equation}
where we have set $\Delta\equiv B\frac{\tan^{-1}( r\gamma)}{r\gamma}$.
The coefficients $C$, $A$, $B$ and $\gamma$ are defined as follows
\begin{eqnarray}\label{a2}
&A=-\frac{\eta}{3\beta}\,,\,\,\, B=\frac{2(1+\gamma^{2})\epsilon}{\zeta+y}\,,\,\,\, \epsilon\ll\mid y-1 \mid\nonumber\\
&\gamma=\sqrt{\frac{\eta}{\mid \beta\mid}}\frac{\zeta
+y}{\zeta-3y}\,,\,\,\, C=1-\frac{2\epsilon}{\zeta+y}\,,\,\,\,\,  y=\frac{\Lambda
\beta}{\eta}\,.
\end{eqnarray}
Here $\epsilon$ is a small parameter which marks the deviation from the inherent de Sitter solution.
Also the scalar field in this case is given by
\begin{equation}
\psi^{2}(r)=-\frac{2(2\eta)^{2}r^{2}(2\eta+2\Lambda\beta)[(2\eta-2\Lambda\beta)r^{2}+4\beta]^{2}}{2\beta(2\eta-2\Lambda\beta)^{2}(2\eta r^{2}+2\beta)^{3}f(r)}\,,
\end{equation}
where $\psi=\frac{d\phi}{dr}$. As we have said, singular points can be obtained by solving the
relation $f(r) = 0$. From this relation singular point is located at
$r_{sing}=\frac{\mu}{B+C}$.

An important issue should be stressed here: about the behavior of the scalar field at the black hole horizon (say, $r_{h}$), the scalar field seems to be divergent at the horizon from Eq. (71). But since $f(r_h )=0$ and $\frac{df}{dr}(r_h)\neq 0$, one can expand $f(r)$ in Taylor series to find $f(r)=f_0+f_{1}(r-r_{h} )+f_{2}(r-r_{h})^{2}+...$ .
For $\psi(r)$ this approximates to $\frac{1}{\sqrt{f(r)}}$ in near horizon which gives for the scalar field $\phi(r)=\phi_{0}+\phi_{1}(r-r_h )^{1/2}+\phi_{2}(r-r_h )^{3/2}+...$ . Therefore, the scalar field remains finite in the near horizon region (a similar analysis can be found in the paper by Miao and Xu (2016)). We note that although the scalar field itself does not diverge at the horizon, its derivative does. However, there is no physical divergence since all
invariants, such as $g^{\mu\nu}\partial_{\mu}\phi \partial_{\nu}\phi$ remain finite. For more discussion on this issue we refer to Feng et al. (2015) (see also Hadar and Reall (2017) and Caceres et al. (2017)).

We note that for nonzero $\eta$ the model given by Eqs. (68)-(70) admits solutions in
which the $\Lambda$-term in action (1) is totally screened. The metric then \emph{is not
asymptotically flat} but rather it is de Sitter with the effective cosmological constant proportional to $\eta$/$\beta$
since the scalar kinetic term tends to a constant around the present time (Gubitosi and Linder, 2011).
It is important to note that as has been shown by Tretyakova (2016), for $1-(B+C) < 3 \times 10^{-4}$
this metric must be equivalent to the Schwarzschild metric in the sense that
it matches with the observations of the gravitational light deflection and perihelion precession.
This feature guarantees the existence of bounds orbits for $(B+C)>0$ (for more discussion on this issue see Tretyakova (2016)).

It is necessary to mention that for calculations, the assumption
$A\simeq 0$ is considered. Then the components of metric (2) will be
as $h(r)=C-\frac{\mu}{r}+\Delta$ and $f(r)=\frac{h(r)}{B+C}$. On the
other hand, since $\gamma\simeq \sqrt{A}$, due to smallness of
$\gamma$, it is more suitable to substitute $\tan^{-1}(r\gamma)\simeq
r\gamma$ and therefore $\Delta\simeq B$. All calculations are done
with these approximations and also with assumption $C=1$
and $0<B+C<1.125$ (Tretyakova 2016). By these assumptions, we
focus mainly on the role of parameter $B$ as Horndeski/Galileon correction factor
in our forthcoming treatment.

\subsection{Circular Equatorial Geodesics}

In order to  investigate circular geodesics in equatorial plane, we
need the explicit form  of the effective potential which is governed
by  equation (12) as
\begin{eqnarray}\label{a17}
V_{eff}=\Big(C-\frac{\mu}{r}+Ar^{2}+\Delta\Big)(1+\frac{L^{2}}{r^{2}})\,,
\end{eqnarray}
where $\Delta\equiv B\frac{\tan^{-1} (r\gamma)}{r\gamma}$. From the condition $\frac{d^{2}}{dr^{2}}V_{eff}>0$ for existence of the
stable circular orbits, we see that for
$r<\frac{3}{2}\frac{\mu}{B+C}$ and $r>3\frac{\mu}{B+C}$ this
condition  holds, which with regard to equation (20), the location
of the stable circular orbits would be at $r\geq 3\frac{\mu}{B+C}$.
Then
\begin{eqnarray}\label{a18}
r_{isco}= 3\frac{\mu}{B+C}\,,
\end{eqnarray}
is introduced as the radius of the innermost stable circular orbit.
The left panel of figure 1 represents the effective potential
versus $r$ for several values of the angular momentum $L$ in the
case with $B=-0.1$. We see that for $L<2.2 \sqrt{3}$ no extremum
can be observed and the first extremum is observed at $L=2.2 \sqrt{3}$
(solid circle in the figure).  This point represents the location of
the innermost stable circular orbit located at $r=6.6$. For larger
values of the angular momentum, $V_{eff}$ has two extremum where the
maximum one denotes the location of the unstable circular orbit and the
minimum one denotes the stable circular orbit. By increasing the
angular momentum, $V_{eff}$ will be larger and the maximum point
turns to the smaller radii whereas the minimum point goes to larger radii.
From the right panel of figure 1 the effect of
Horndeski/Galileon correction factor $B$ on the effective
potential can be seen. The effective potential achieves larger
values for larger values of $B$ and by increasing this parameter,
the loci of unstable circular orbit becomes closer to the central
mass and stable orbits will be located farther from the central mass.
Also the enhancement of the distance between these points by
increasing $B$ is obvious.

\begin{figure}\label{a1}
\begin{center}
\includegraphics{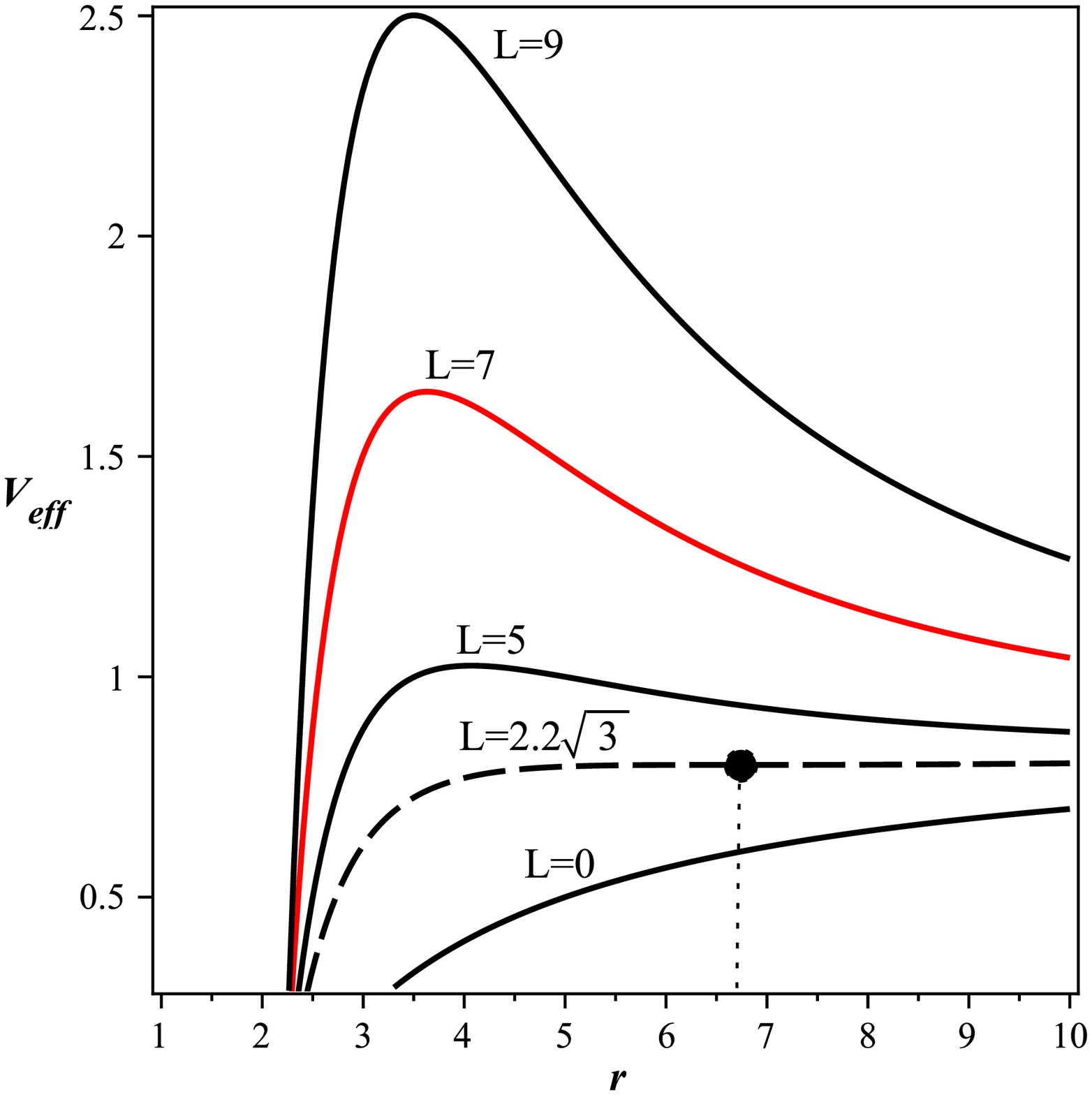}
 \includegraphics{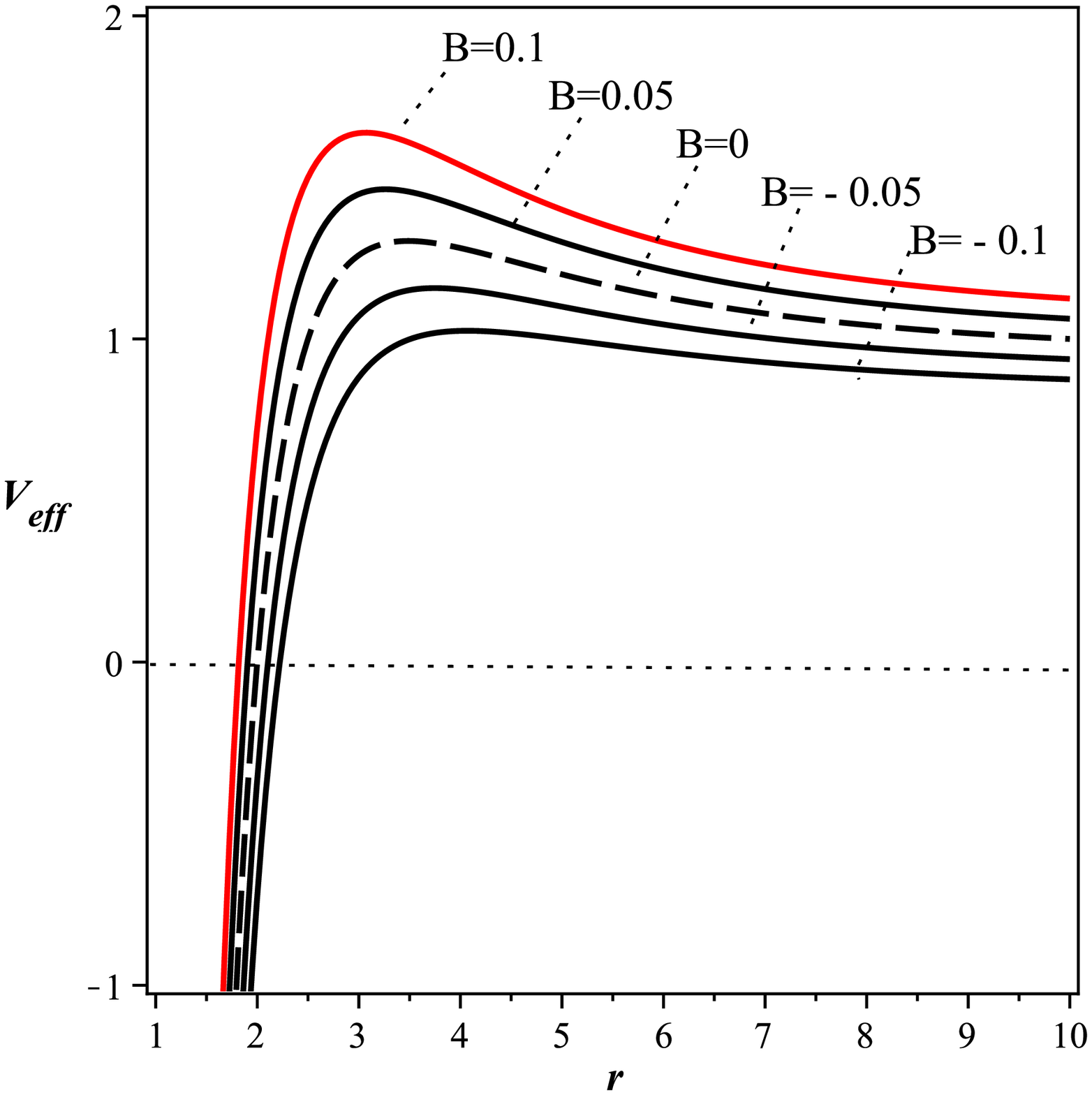}
\end{center}
\vspace{6cm} \caption{Effective potential for massive particles
versus the radial coordinate (in the unit of $\frac{1}{r_{g}}$). The
left panel represents $V_{eff}$ for $B=-0.1$ and for several values of
the angular momentum. The solid circle in the left panel denotes the effective
potential in ISCO. In the right panel, effect of the Horndeski/Galileon correction parameter $B$
on effective potential for $L=5$ is shown. }
\end{figure}

In addition to innermost stable circular  orbit which is very
important in studying the accretion around the black hole, there are
other special radii where considering them is necessary. As we have
stated previously, the circular orbits exist only for radii larger
than the photon radius $r_{ph}$. For $r_{ph}<r<r_{ms}$, the motion of
the particle will be unstable against the small perturbations. This
means that particle falls into the black hole or flee away to
infinity. In the region $r>r_{ms}$ the particle moves on stable
circular orbits.

From equations (18)-(20), other characteristic radii including
the photon sphere $r_{ph}$, circular orbit $r_{circ}$ and
marginally bound orbit $r_{mb}$ can be obtained respectively as
\begin{eqnarray}\label{a18}
r_{ph}=\frac{3}{2}\frac{\mu}{B+C}\,,
\end{eqnarray}
\begin{eqnarray}\label{a18}
r_{circ}>\frac{3}{2}\frac{\mu}{B+C}\,,
\end{eqnarray}
and
\begin{eqnarray}\label{a18}
r_{mb}=\frac{1}{4}\mu \frac{4(B+C)-3\pm
\sqrt{9-8(B+C)}}{(B+C)(B+C-1)}\,.
\end{eqnarray}

We have plotted the characteristic radii versus the
Horndeski/Galileon correction factor $B$ in figure 2. The value of
this parameter affects the location of the characteristic radii in
the vicinity of the black hole. For larger values of $B$, the
location of $r_{ph}$, $r_{isco}$ and $r_{sing}$ will be closer to
the black hole, whereas the behavior of $r_{mb}$ for negative and
positive values of $B=0$ is different. For $B<0$, marginally bound
radius decreases by increasing $B$, but for $B>0$ this behavior
becomes reverse as it finally matches the innermost stable circular
orbit. Since $r_{isco}$ represents the inner edge of the accretion
disk, we see that in larger values of $B$, the disk will be extended
close to the central mass.


\begin{figure}\label{a2}
\begin{center}
\includegraphics{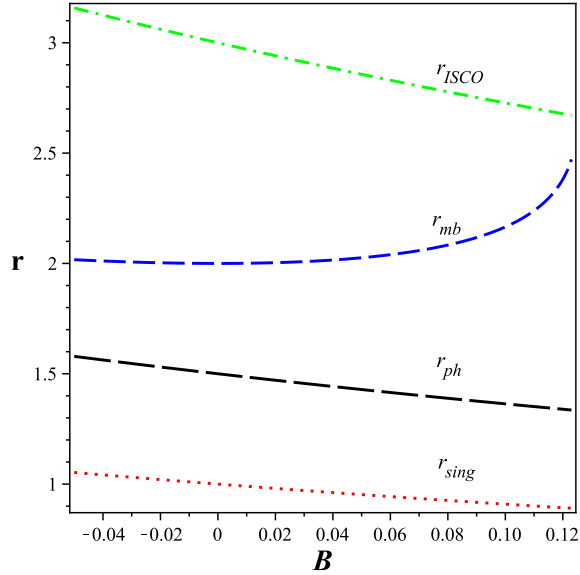}
\end{center}
\vspace{7cm} \caption{The effect of the Horndeski/Galileon correction factor $B$ on characteristic
radius and comparison between these radii. The vertically dotted line
represents the location of these radii in Schwarzschild spacetime.}
\end{figure}

For a particle which moves in a circular orbit, the specific energy,
specific angular momentum, angular velocity and angular momentum in
equatorial plane can be derived as
\begin{eqnarray}\label{a18}
E^{2}=\frac{2\Big(C-\frac{\mu}{r}+Ar^{2}+\Delta\Big)^{2}}{\Big(2C-3\frac{\mu}{r}+3\Delta-\frac{B}{1+r^{2}\gamma^{2}}\Big)}\,,
\end{eqnarray}
\begin{eqnarray}\label{a17}
L^{2}=r^{2}\frac{\Big(\frac{\mu}{r}+2Ar^{2}-\Delta+\frac{B}{1+r^{2}\gamma^{2}}\Big)}{\Big(2C-3\frac{\mu}{r}+3\Delta-\frac{B}{1+r^{2}\gamma^{2}}\Big)}\,,
\end{eqnarray}
\begin{eqnarray}\label{a18}
\Omega_{\varphi}^{2}=\frac{1}{2r^{2}}\Big(\frac{\mu}{r}+2Ar^{2}-\Delta+\frac{B}{1+r^{2}\gamma^{2}}\Big)\,,
\end{eqnarray}
\begin{eqnarray}\label{a18}
l^{2}=\frac{r^{2}}{2}\frac{\Big(\frac{\mu}{r}+2Ar^{2}-\Delta+\frac{B}{1+r^{2}\gamma^{2}}\Big)}{\Big(-\frac{\mu}{r}+Ar^{2}+\Delta+C\Big)^{2}}\,,
\end{eqnarray}
respectively. For the innermost stable circular orbit, these relations
reduce to
\begin{eqnarray}\label{a18}
E^{2}_{isco}= \frac{8}{9}(B+C)\,,
\end{eqnarray}
\begin{eqnarray}\label{a18}
L^{2}_{isco}= \frac{3\mu ^{2}}{(B+C)^{2}}\,,
\end{eqnarray}
\begin{eqnarray}\label{a18}
\Omega^{2}_{isco}=\frac{1}{54}\frac{(B+C)^{3}}{\mu^{2}}\,,
\end{eqnarray}
\begin{eqnarray}\label{a18}
l^{2}=\frac{27}{8}\frac{\mu^{2}}{(B+C)}\,,
\end{eqnarray}
respectively.

\begin{figure}\label{a3}
\begin{center}
\includegraphics{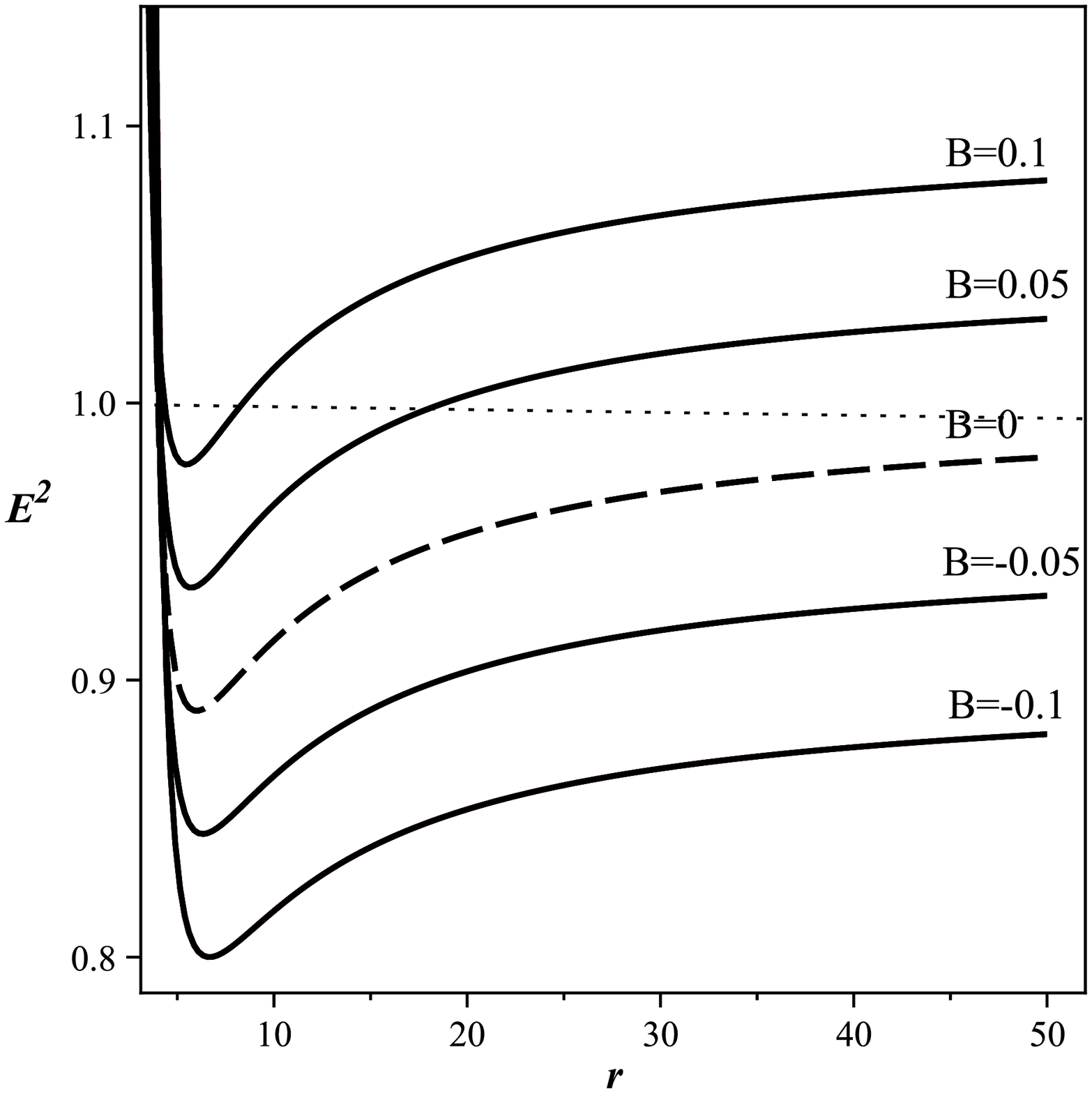}
 \includegraphics{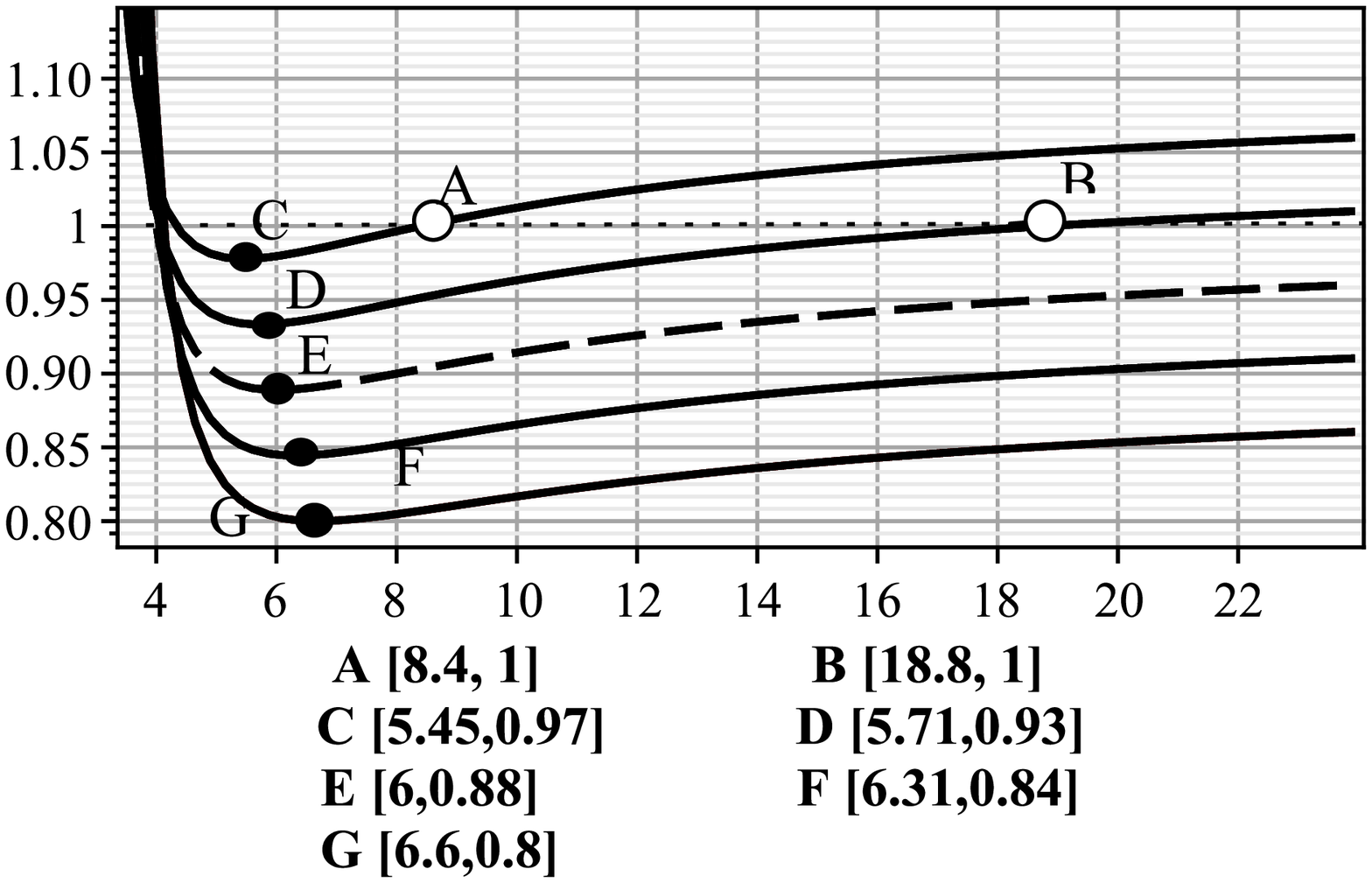}
 \includegraphics{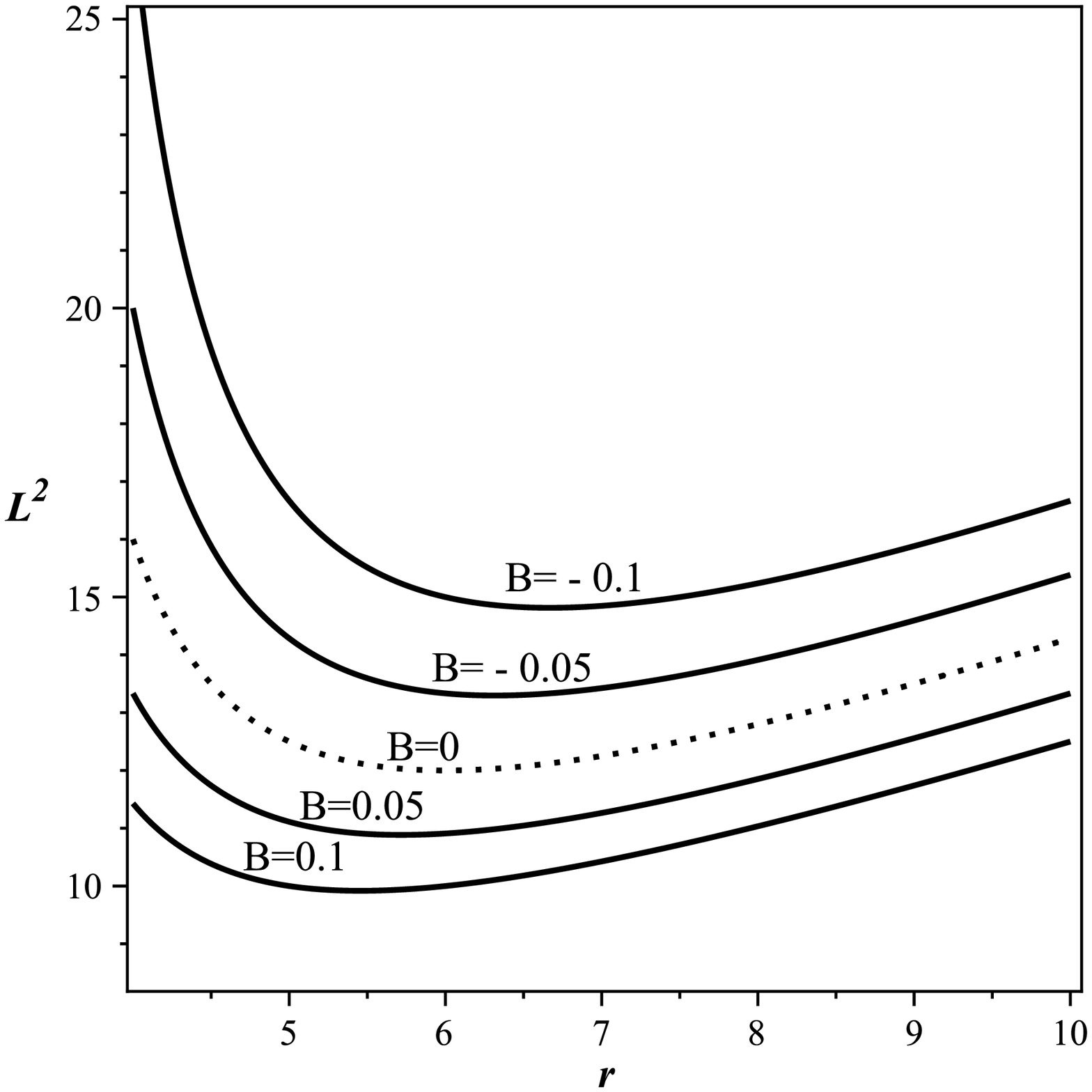}
 \includegraphics{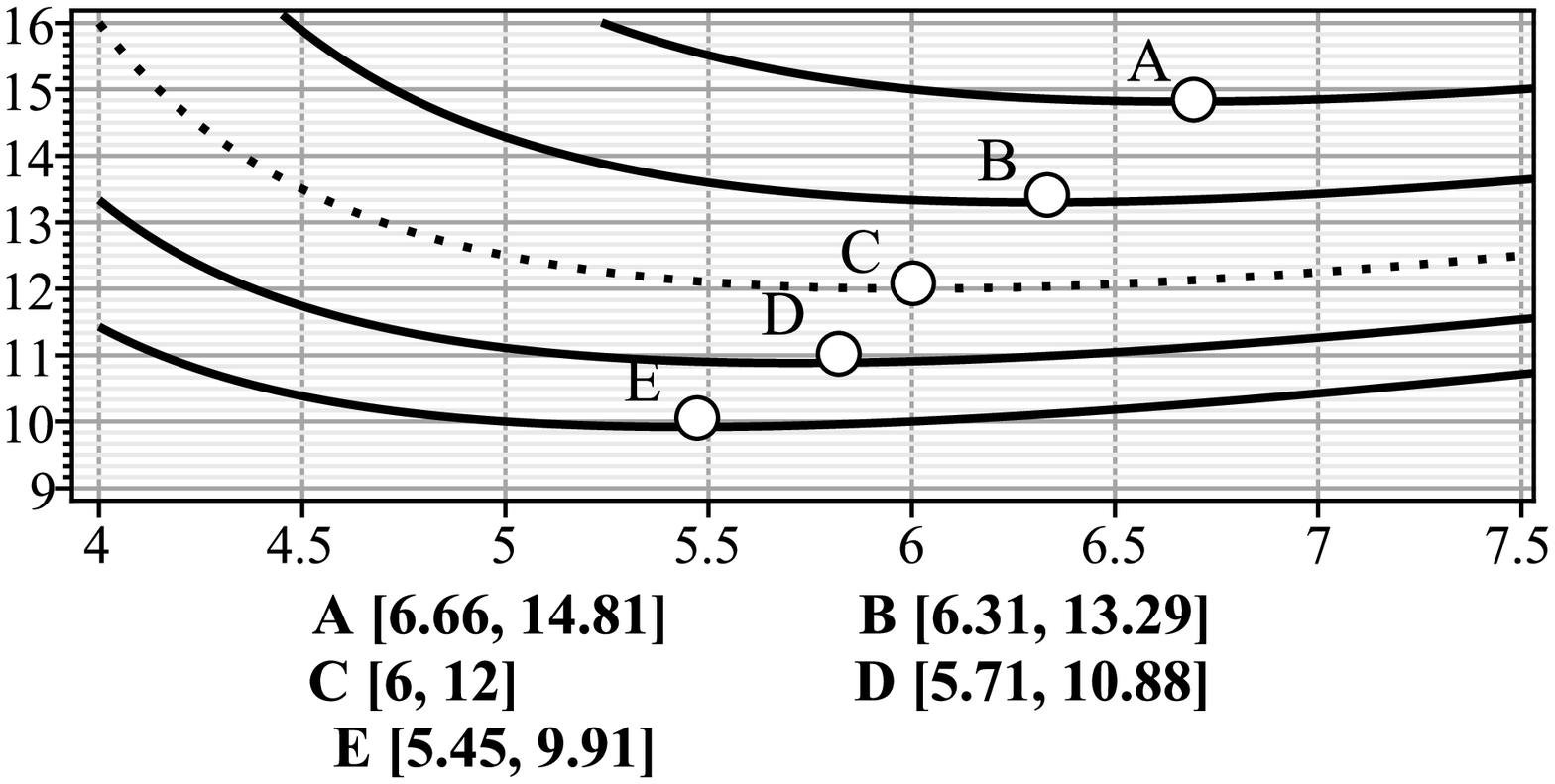}
\end{center}
\vspace{15cm} \caption{The behavior of energy (upper panel) and
angular momentum (lower panel) versus the radial distance from the
central mass for several values of the Horndeski/Galileon parameter $B$.
The innermost stable circular orbits are represented by the solid circles in the upper right panel. The empty circles
at the energy diagram denote the limit of the bound orbit. The lower panel shows that the angular
momentum gets smaller values by increasing $B$. The loci of these
parameters in innermost stable circular orbits are represented by
circles on the lower right panel of the figure.}
\end{figure}

\begin{figure}\label{a4}
\begin{center}
\includegraphics{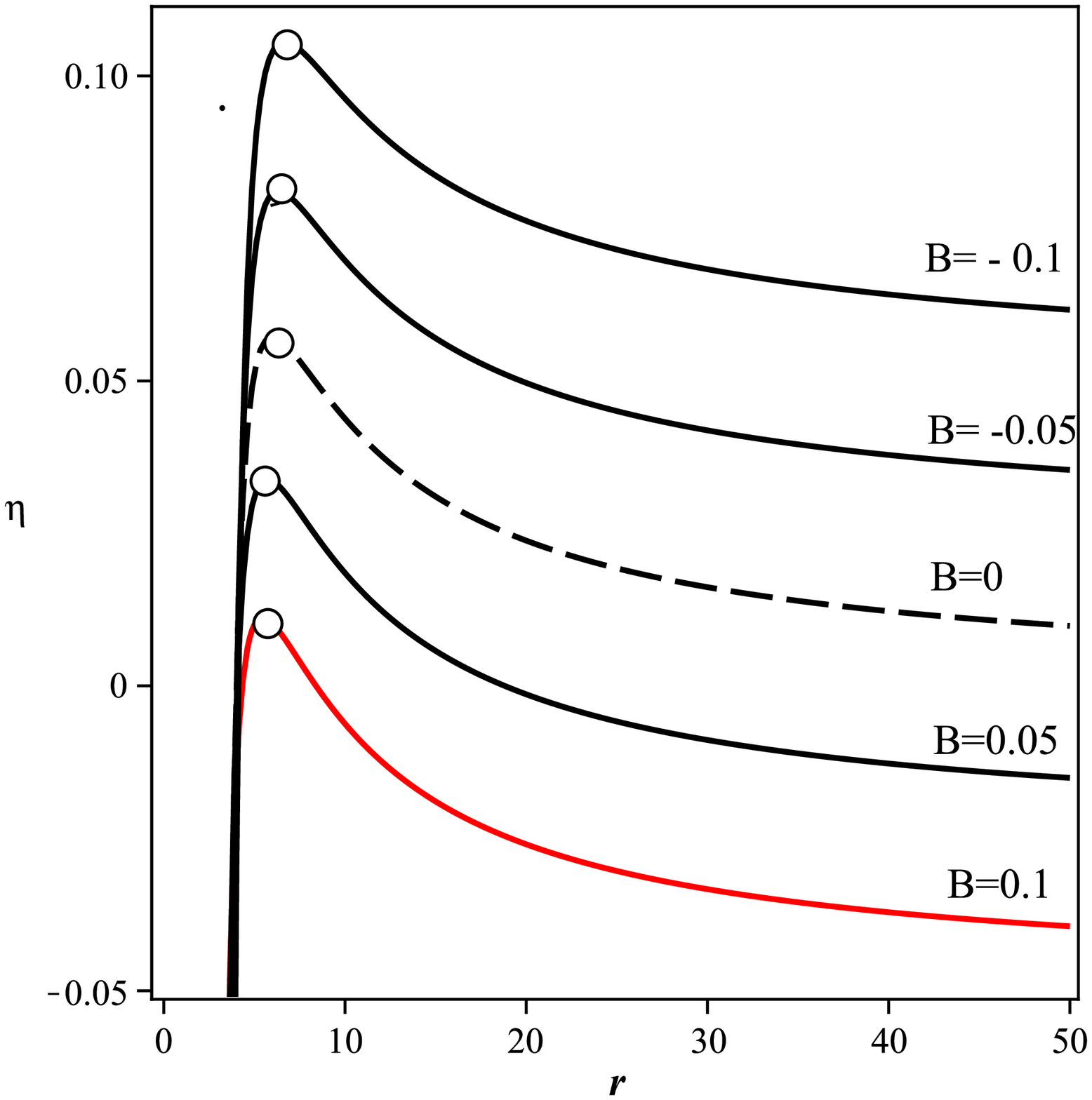} \includegraphics{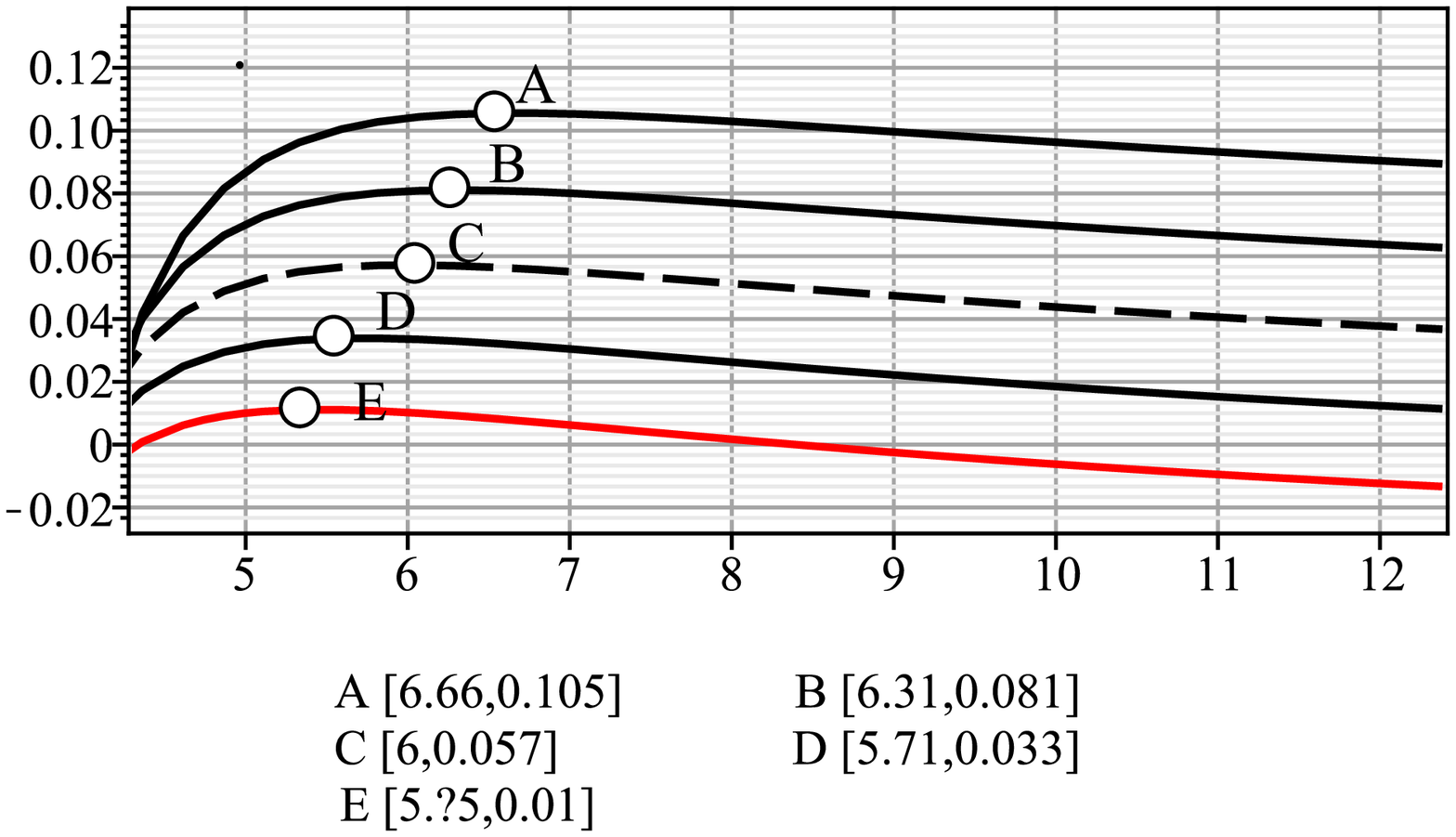}
\end{center}
\vspace{6cm} \caption{Energy efficiency of a massive particle
falling from infinity into the black hole is shown versus the radial
distance for several values of the Horndeski/Galileon parameter $B$. The maximum
efficiency at the innermost stable circular orbit is represented by a
circle in each case in the right panel.}
\end{figure}


\begin{figure}\label{a5}
\begin{center}
\includegraphics{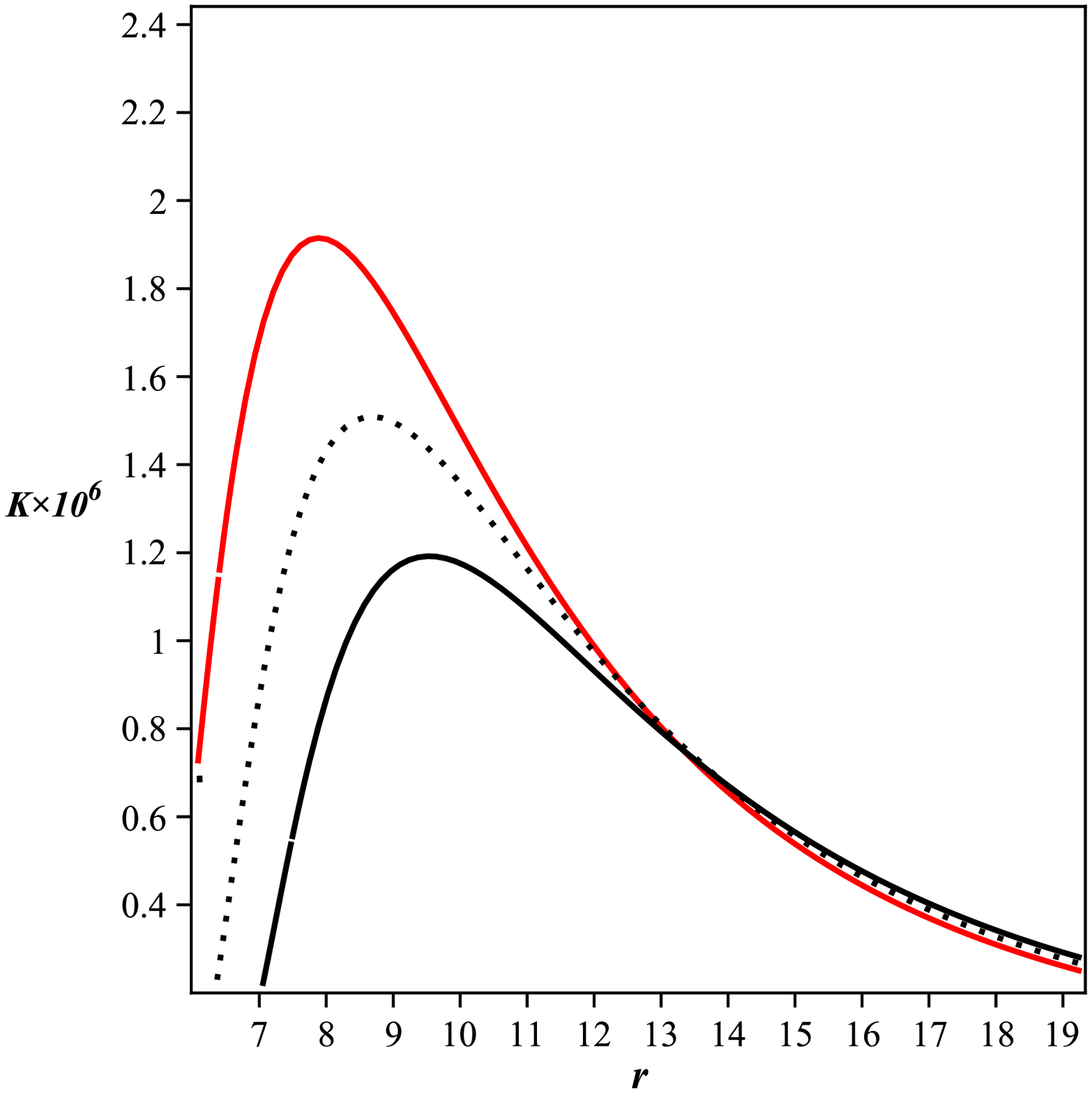}
 \includegraphics{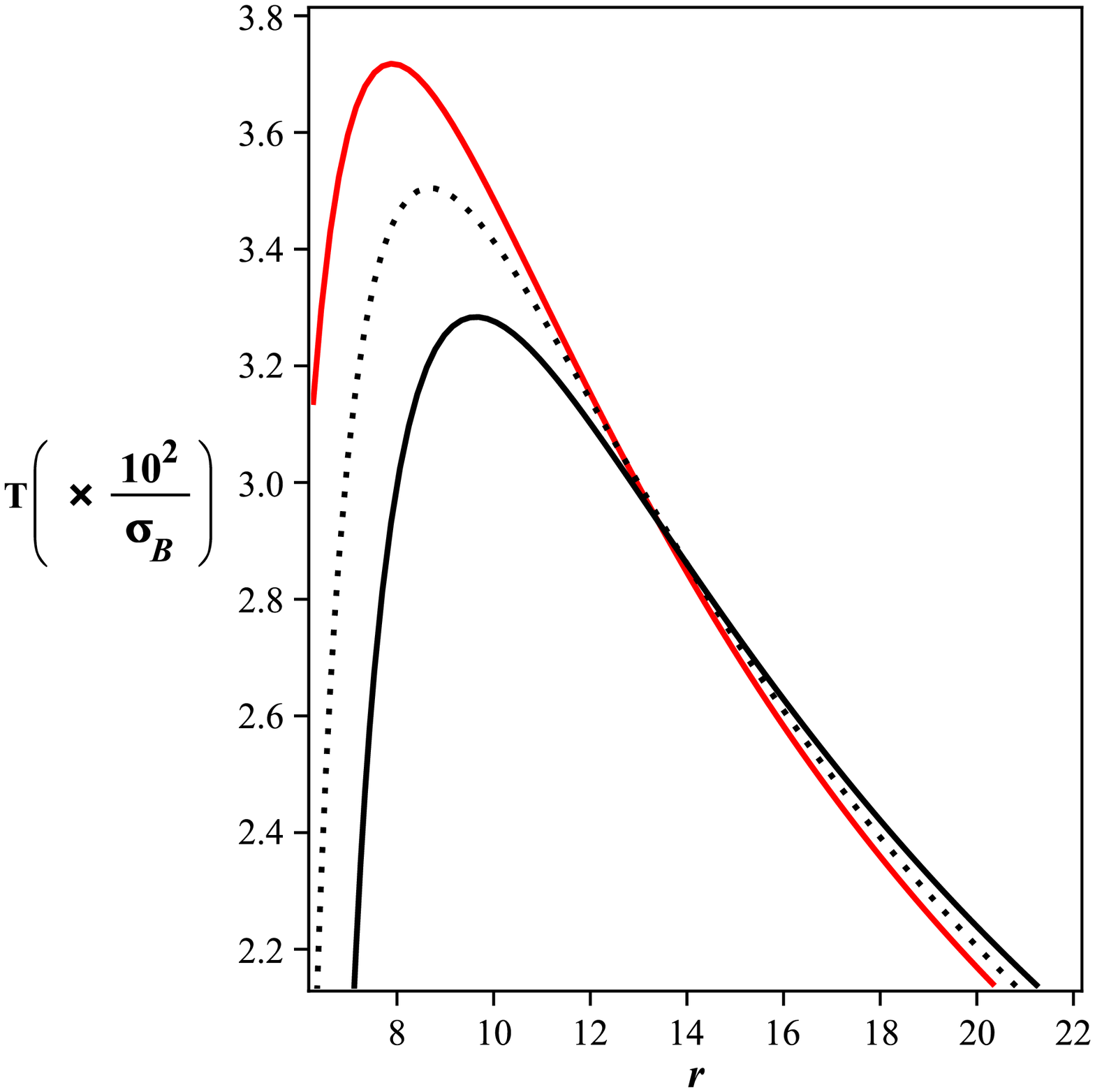}
\end{center}
\vspace{6cm} \caption{Dependence of the emission rate and
temperature $T$ on radius $r$ and Horndeski/Galileon parameter $B$ ($B=-0.1, 0,  0.1$
from top to down respectively). Dotted curve represents the case of the
Schwarzschild black hole ($B=0$).}
\end{figure}

\begin{figure}\label{a6}
\begin{center}
\includegraphics{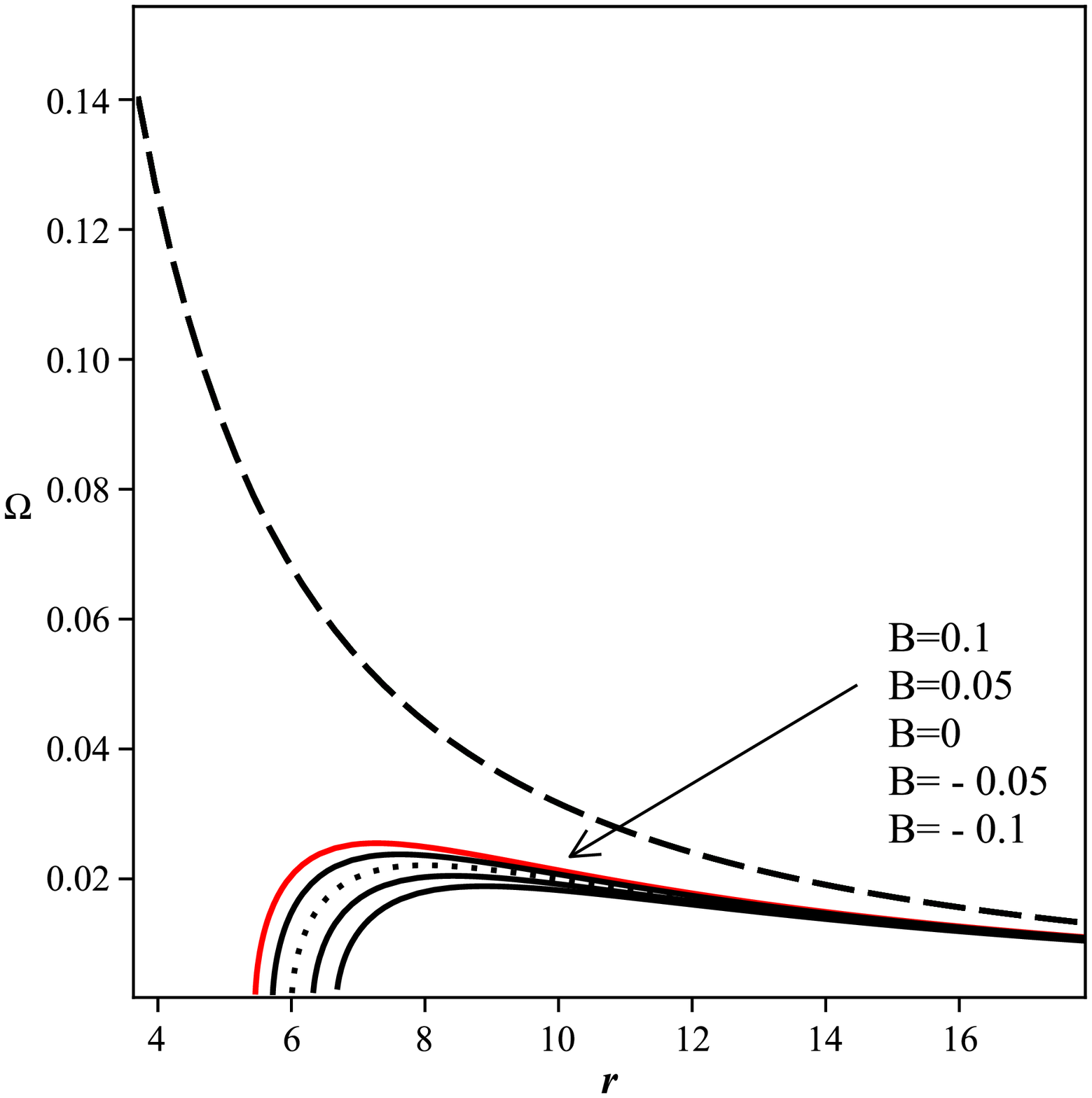}
 \includegraphics{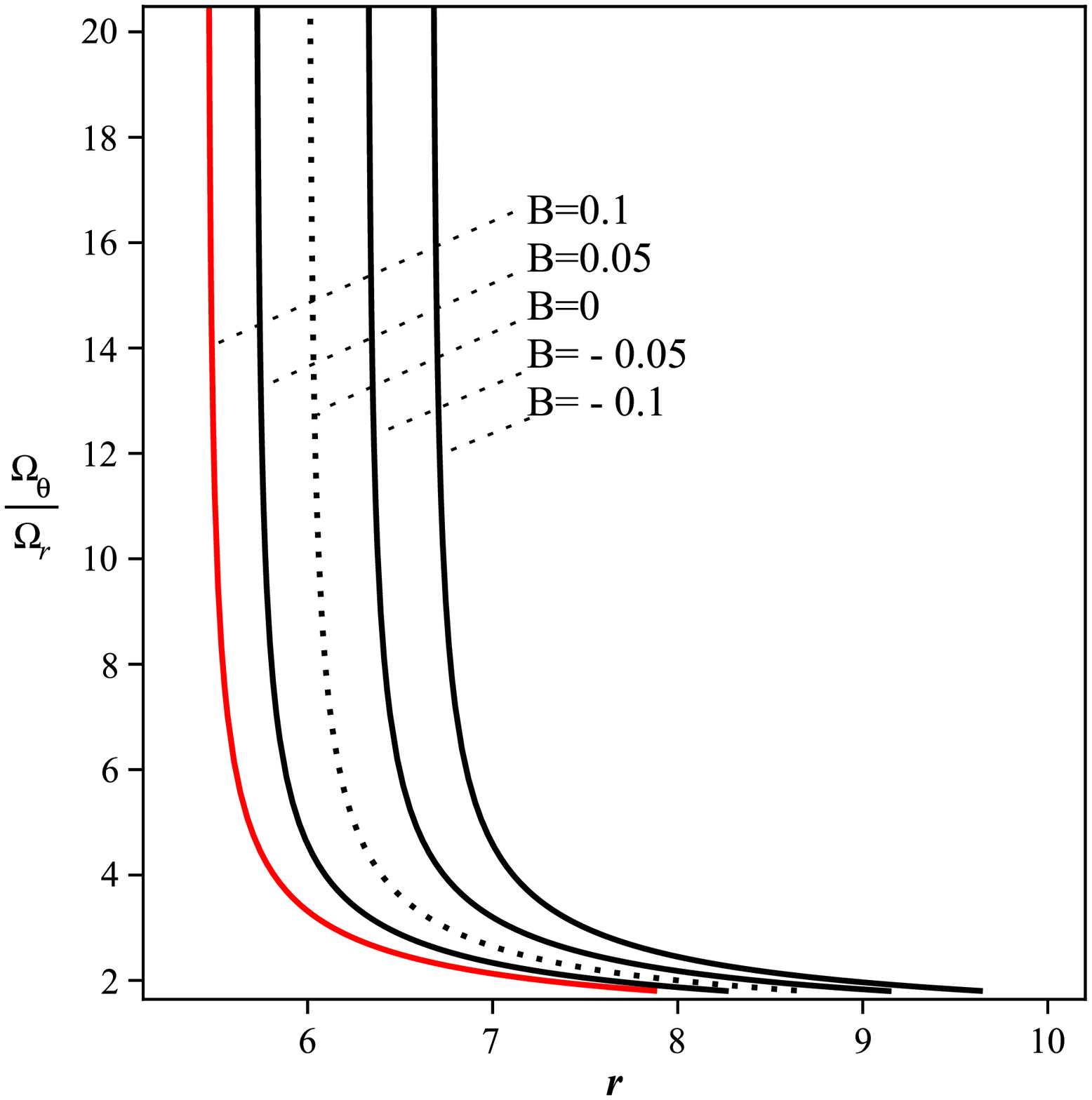}
\end{center}
\vspace{6cm} \caption{A comparison between the epicyclic frequencies.
The dashed curve in the left panel represents the vertical
frequency. The radial epicyclic frequencies for several values
of the Horndeski/Galileon parameter $B$ are shown by the solid curves where the dotted curve is for the Schwarzschild
black hole with $B=0$. The effect of Horndeski/Galileon parameter on the ratio of the vertical
frequency to the radial frequency as a function of $r$ is shown in the
right panel. }
\end{figure}

\begin{figure}\label{a7}
\begin{center}
\includegraphics{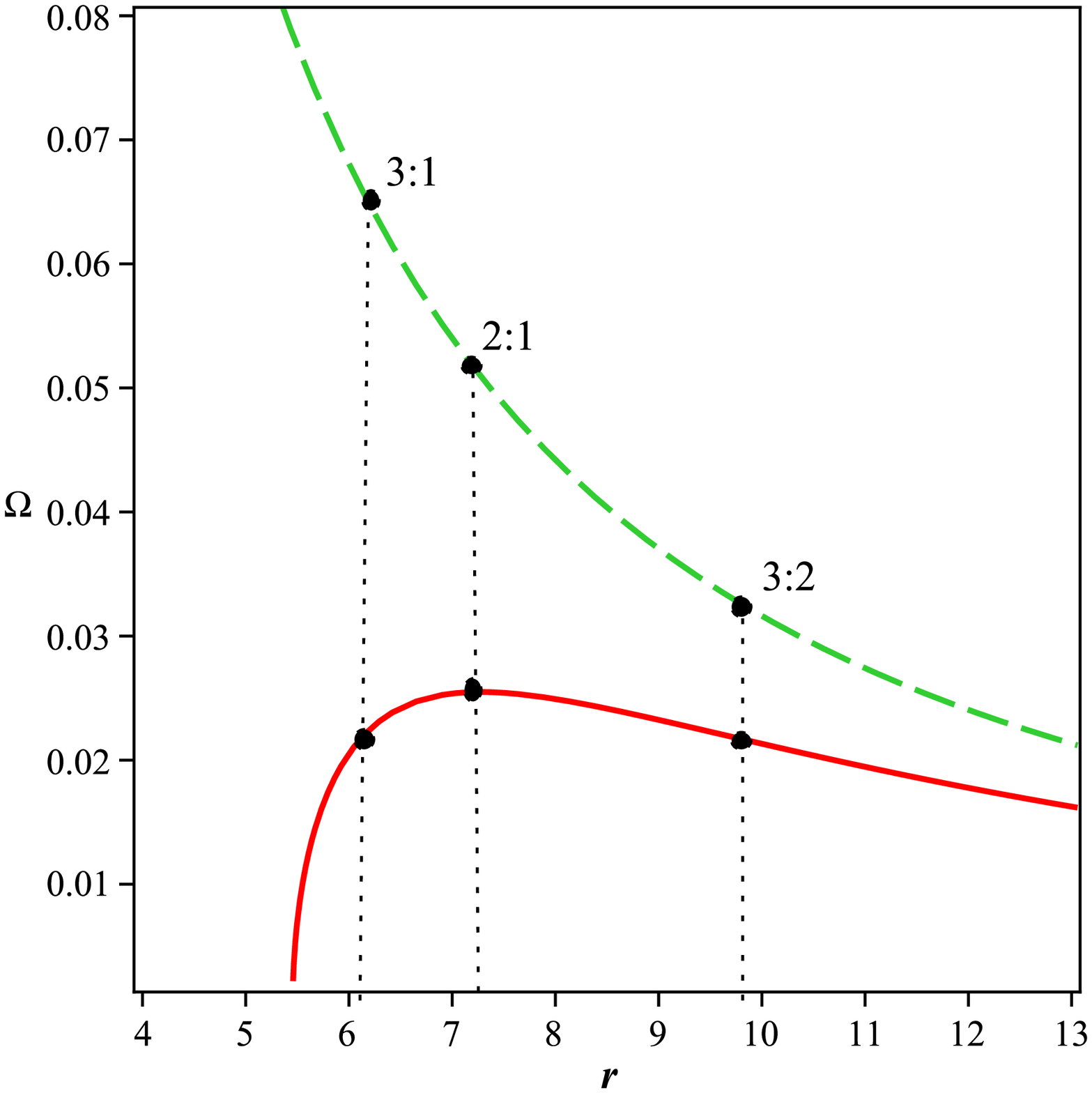}
 \includegraphics{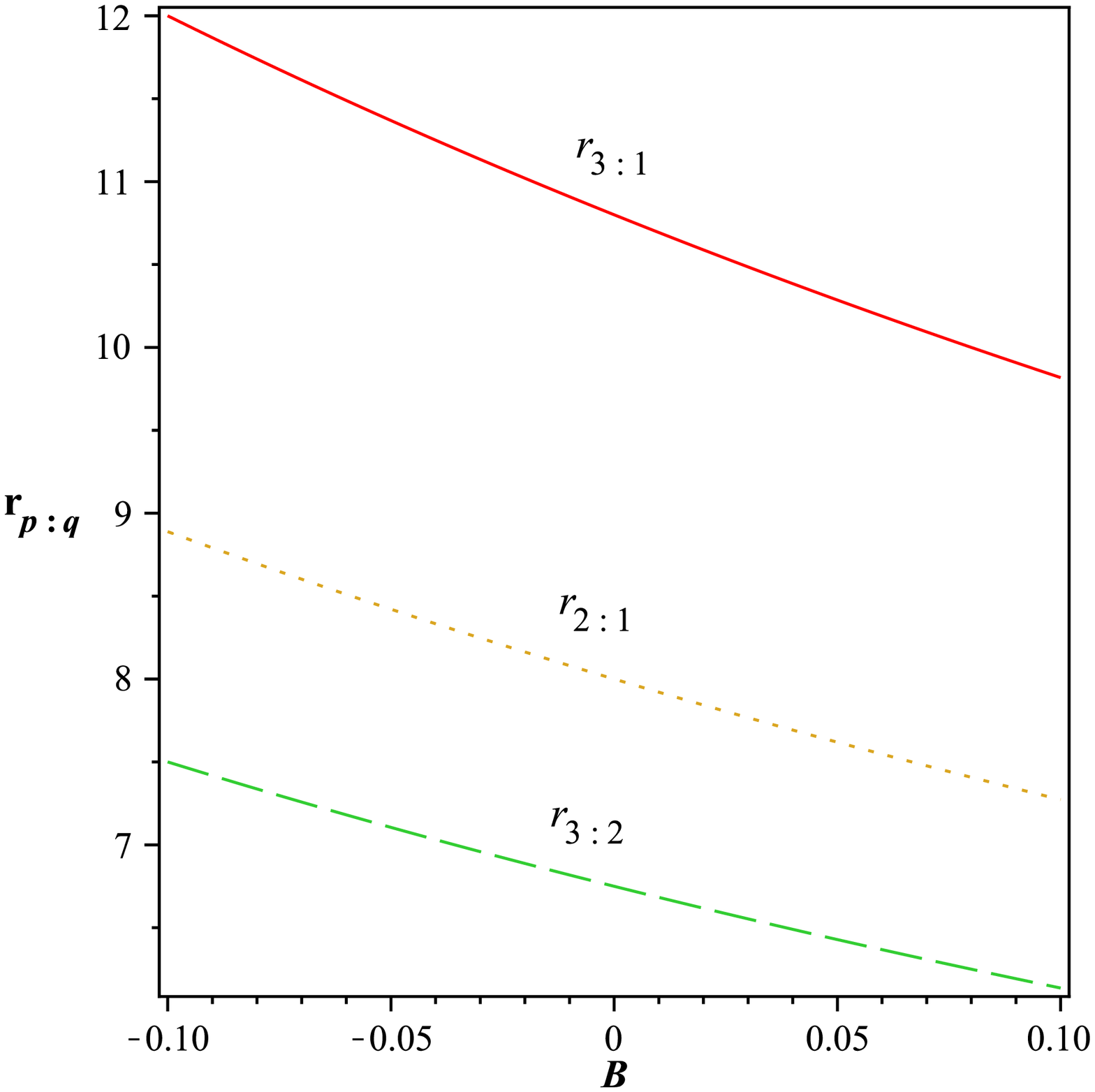}
\end{center}
\vspace{7cm} \caption{ The left panel represents the location of three
resonances: parametric resonance(3:2) and forced resonance (3:1, 2:1).
The dependence of these locations to the Horndeski/Galileon parameter $B$ is shown
in the right panel. }
\end{figure}

\begin{figure}\label{a7}
\begin{center}
\includegraphics{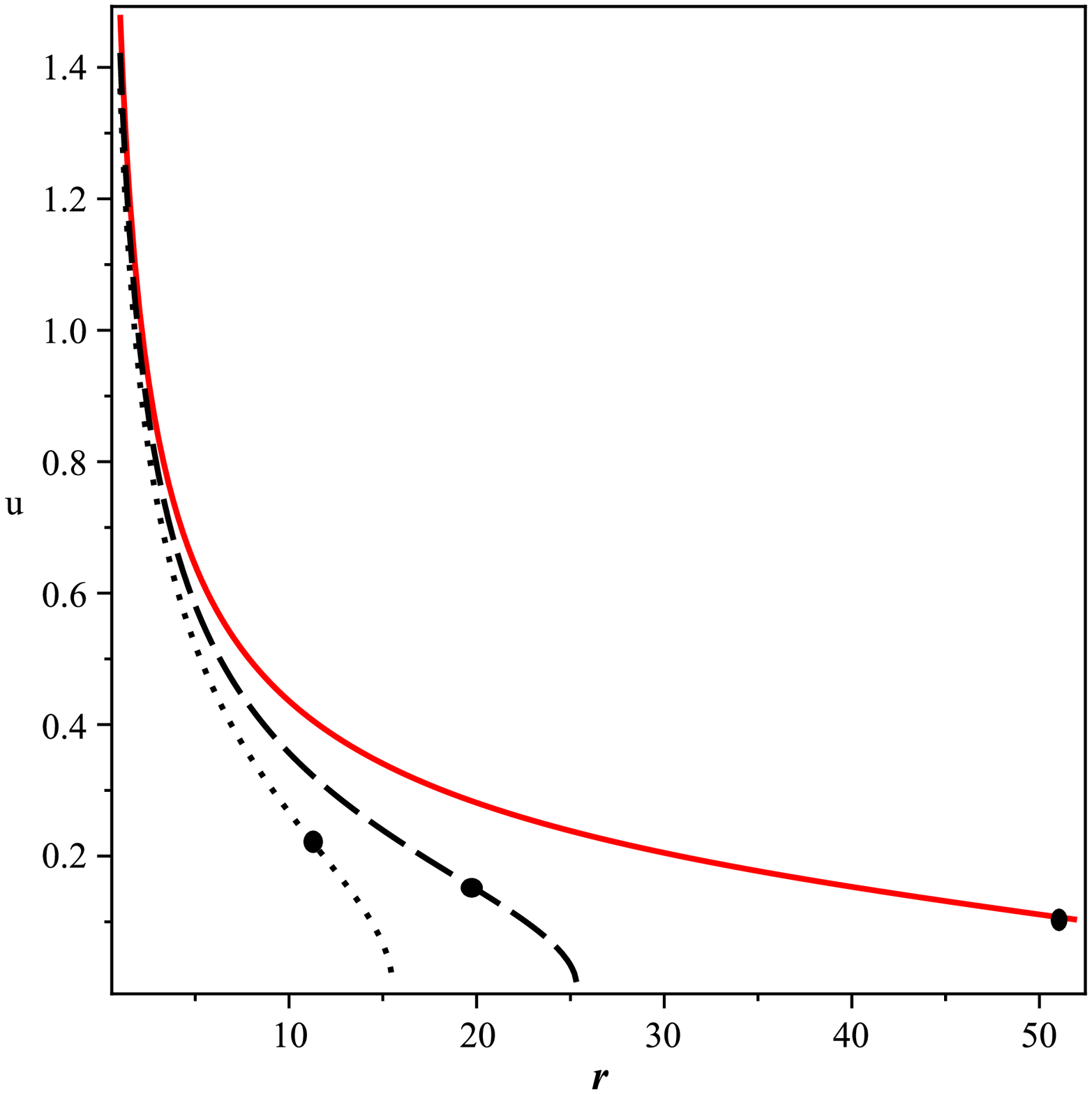}
 \includegraphics{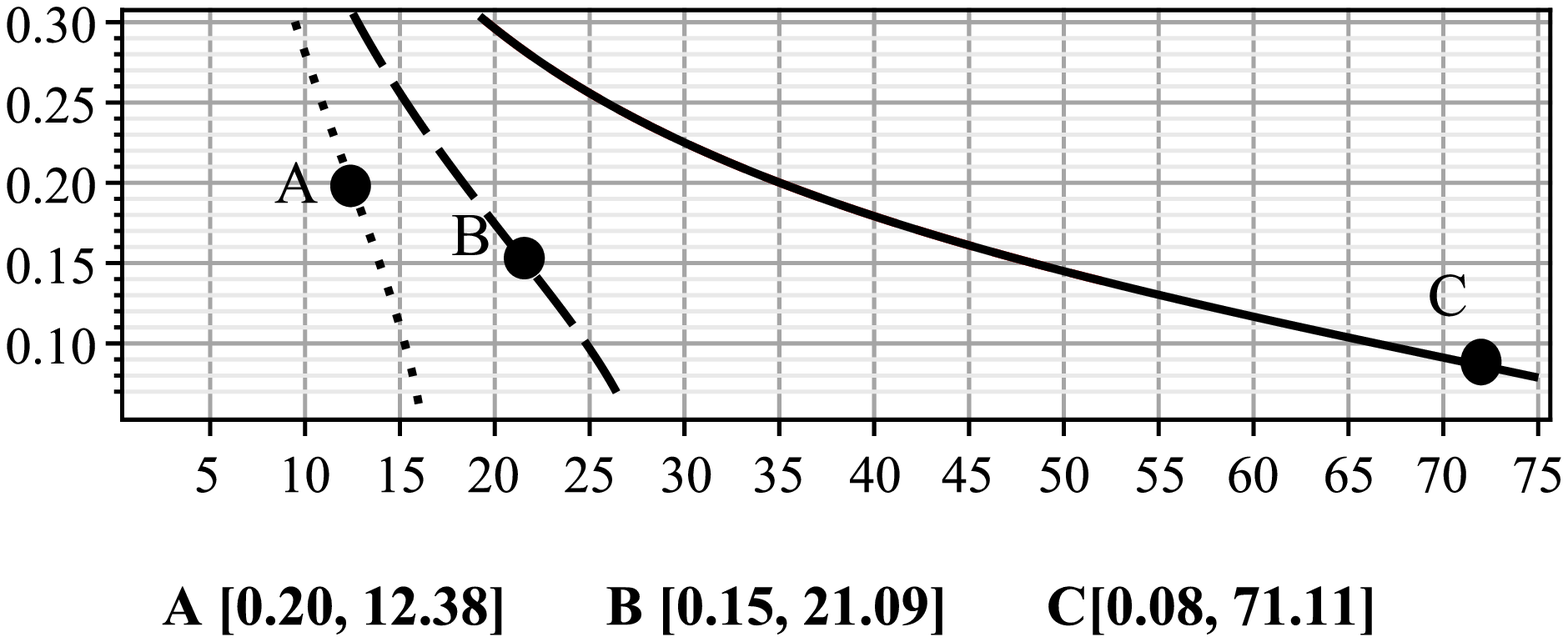}
\end{center}
\vspace{7cm} \caption{ The radial velocity profile as a function of
the dimensionless parameter $r$ for the equation of state parameter
$k=\frac{1}{2}$ and constant of integration $A_{4}=1.4$. The dotted,
dashed and solid lines are corresponding to the cases with $B=0$ (Schwarzschild geometry), $B=-0.05$ and
$B=-0.1$ respectively. For each curve, the critical radii are marked
by a solid circle which their coordinates are stated in the right
panel. }
\end{figure}

In figure 3, the behavior of specific energy and angular momentum
versus the radius are shown and the effect of the Horndeski/Galileon
correction factor $B$ is studied. The upper left panel represents
variation of the specific energy versus $r$. In the upper right panel
the location of this radius in innermost stable circular orbits is
shown by the solid circle. The empty circle denotes the limit of
the bound orbit. Increasing the Horndeski/Galileon correction
factor enhances the energy and decreases the range of bound
orbit radius. The lower left panel represents the angular momentum where
it gets smaller values by increasing $B$. The loci of these
parameters in innermost stable circular orbits are represented by
circles on the lower right panel of the figure.

Now by knowing $E$ in ISCO, we are able to determine the radiation
energy efficiency of accretion. As we have said previously,
the efficiency of accretion is $1-E$, where the maximum efficiency of
accretion given by $1-E_{isco}$ in innermost stable orbit in this setup is
$1-\sqrt{\frac{8}{9}(B+C)}$. In figure 4, the efficiency of
accretion is plotted versus $r$ for different values of
the Horndeski/Galileon correction factor $B$ where we
have denoted the maximum with a circle in this figure. In the case $B=0$
where coincides with the Schwarzschild black hole (dashed line),
the efficiency equals to $0.057$ at $r=6 r_{g}$ as usual. For negative values
of the Horndeski/Galileon parameter, efficiency goes up and becomes greater than the
Schwarzschild black holes ones. For example, in the case with
$B=-0.1$, the efficiency equals to $0.105$ at $r=6.66 r_{g}$ where
is tangible. On the other hand, positive values of $B$ is
accompanied with decreasing of the efficiency.

Now we can study radiation flux from the surface of the accretion
disk by knowing $E$, $L$ and $\Omega_{\varphi}$ in equatorial plane
and $r_{ms}$. The Flux of the radiation energy of the accretion disk
is obtained from equations (24) and (25) as
\begin{eqnarray}\label{a18}
&K(r)=-\frac{1}{8}\dot{M}\sqrt{\frac{2r^{2}(1+\frac{\eta}{\beta}r^{2})}{(3Ar^{2}+C+\frac{B}{1+r^{2}\gamma^{2}})(2Ar^{2}+\frac{\mu}{r}+
\frac{B}{1+r^{2}\gamma^{2}}-\Delta)}} \times \nonumber\\
&\frac{1}{r}\Big\{3\Delta+2C-3\frac{\mu}{r}-\frac{B}{1+r^{2}\gamma^{2}}\Big\}
\Big\{3\Delta-3\frac{\mu}{r}-3\frac{B}{1+r^{2}\gamma^{2}}-2\frac{Br^{2}\gamma^{2}}{(1+r^{2}\gamma^{2})^{2}}\Big\}\times \nonumber\\
&\Big\{\pi
r^{4}(\Delta+4Ar^{2}+2C-\frac{\mu}{r}+\frac{B}{1+r^{2}\gamma^{2}})^{2}\Big\}^{-1}\int_{r_{_{m_{s}}}}^{r}\mathcal{F}(r)
dr\,,
\end{eqnarray}
where by definition
\begin{eqnarray}\label{a18}
&\mathcal{F}(r)=\frac{1}{2}\Big\{\sqrt{\frac{2r^{2}}{2Ar^{2}+\frac{\mu}{r}-\Delta+\frac{B}{1+r^{2}\gamma^{2}}}}(4Ar^{2}-\frac{\mu}{r}+
\Delta+\frac{B}{1+r^{2}\gamma^{2}}+2C)\times\nonumber\\
&\Big(\frac{1}{r}(\Delta+Ar^{2}+C-\frac{\mu}{r})(\Delta-8Ar^{2}-\frac{\mu}{r}-\frac{B}{1+r^{2}\gamma^{2}}+\frac{2Br^{2}\gamma^{2}}{(1+r^{2}\gamma^{2})^{2}})\times\nonumber\\
&+2(\frac{B}{1+r^{2}\gamma^{2}}-\Delta+2Ar^{2}+\frac{\mu}{r})\Big)\Big\}\Big\{-3\frac{\mu}{r}+3\Delta+2C-\frac{B}{1+r^{2}\gamma^{2}}\Big\}^{-2}\,.
\end{eqnarray}
Then the temperature can be obtained by using the equation $K=\sigma T^{4}$. In
figure 5, the relation between the radiation flux and temperature is
shown. The radiation flux has a maximum in the vicinity of the black
hole and decreases at the smaller radii. By increasing the Horndeski/Galileon correction factor
$B$, the energy flux raises and the maximum of emission flux tends to the
smaller radii but the reverse happens after this point. The
dependence on  this parameter is very considerable in the vicinity
of the black hole, but it is relatively weak far from the black
hole. These behaviors are the same for the temperature. As we have said in subsection 3.2, the luminosity
can be compute from equation (26), but because of complexity of
the required equations, solving this equation is not possible
analytically.


\subsection{Epicyclic Frequencies}

If a perturbation acts on a particle moving on a circular orbit
in the equatorial plane, the particle experiences small
oscillations in the vertical and radial directions. Using the
epicyclic frequencies given by equations (33) and (34), we derive
the radial and vertical epicyclic frequencies as follows
\begin{eqnarray}\label{a18}
\Omega_{\theta}^{2}=\frac{2Ar^{2}+\frac{\mu}{r}-\Delta+\frac{B}{1+r^{2}\gamma^{2}}}{2r^{2}}\,,
\end{eqnarray}
and
\begin{eqnarray}\label{a18}
&\Omega_{r}^{2}=-\frac{3}{2}(1+\frac{\eta}{\beta}r^{2})\Big\{\Big[
-\frac{8}{9}A\gamma^{4}r^{8}(\frac{15}{8}\Delta+C)+\frac{5}{3}A\gamma^{4}r^{7}\mu+\gamma^{2}
r^{6}\Big(\frac{1}{9}\Delta\gamma^{2}(3\Delta+C)+\nonumber\\
&A(-\frac{16}{9}C-\frac{10}{3}\Delta+B)\Big)+\frac{10}{3}\mu
\gamma^{2}r^{5}\Big((-\frac{1}{30}C-\frac{1}{5}\Delta)\gamma^{2}+\nonumber\\
&A\Big)\Big(\frac{1}{3}\mu^{2}\gamma^{2}+\Big[\frac{1}{9}(2\Delta+B)C-\frac{1}{3}\Delta(-2\Delta+B)\Big]
\gamma^{2}+\frac{7}{9}A(B-\frac{15}{7}\Delta-\frac{8}{7}C)\Big)r^{4}\nonumber\\
&+\Big((\frac{1}{3}B-\frac{2}{9}C-\frac{4}{3}\Delta)\gamma^{2}+\frac{5}{3}A\Big)\mu
r^{3}\Big(\frac{2}{3}\mu^{2}\gamma^{2}+\frac{2}{9}(-\Delta+B)(B-\frac{3}{2}\Delta-\frac{1}{2}C)\Big)r^{2}+\nonumber\\
&\frac{5}{9}(-\frac{6}{5}\Delta-\frac{1}{5}C+B)\mu
r+\frac{1}{3}\mu^{2}\Big]\Big\}\times\nonumber\\
&\Big\{(1+r^{2}\gamma^{2})\Big[Ar^{4}\gamma^{2}+(\frac{1}{3}\gamma^{2}C+A)r^{2}+\frac{1}{3}B+\frac{1}{3}C\Big]r^{4}\Big\}^{-1}\,.
\end{eqnarray}
respectively. From equations (78), (86) and (87) it is clear that
$$\Omega_{\theta}^{2}=\Omega_{\varphi}^{2}$$ and
$$\Omega_{r}^{2}=\Omega_{\theta}^{2}[\frac{C+3\Delta
-2B-3\frac{\mu}{r}}{(B+C)}].$$ In the left hand side of figure 6
we have plotted epicyclic frequencies versus the radius for several
values of the Horndeski/Galileon correction factor $B$ in
order to have a comparison. The angular and vertical epicyclic
frequencies are shown in the left panel of this figure by the
dashed lines which are coincide and will decrease by increasing $r$.  We
see explicitly that they are not dependent on parameter $B$. The
radial epicyclic frequency is shown by the solid and dotted curves
(Schwarzschild black hole) with a maximum where increasing the
parameter $B$ shifts this maximum to the smaller radii. The radial oscillation
frequency increases by increasing the values of the parameter $B$.
Dependence on this parameter is significant close to the black hole,
but far from the central mass this effect is weak. We see that
$\Omega_{r}<\Omega_{\theta}$. The ratio of
$\frac{\Omega_{\theta}}{\Omega_{r}}$ is plotted in the right hand side
panel of figure 6 which is a decreasing function of $r$. It is clear
that in the vicinity of the black hole, this ratio is very greater than
unity but far from the black hole it turns to unity and decreases by
increasing $B$. In the left panel of figure 7, the
locations of three particular resonances such as the parametric
resonance with condition
$\frac{\Omega_{\theta}}{\Omega_{r}}=\frac{3}{2}$ and the forced
resonance with ratio $3:1$ and $2:1$ are shown. The dependence of
these characteristic radii to the metric parameter $B$ is shown in the
right panel of figure 7 where such radii are monotonically
decreasing functions of this parameter. On the other hand, resonance
will be happened in smaller distance from the central mass for larger
values of the metric parameter, $B$.

\subsection{Mass Evolution and Critical Points}

In this Horndeski/Galileon accretion disk, the energy density and radial velocity for an
isothermal fluid are given by
\begin{eqnarray}\label{a18}
\rho=\frac{A_{3}(k+1)}{r^{2}\sqrt{A_{4}^{2}-(k+1)^{2}\Big(\Delta+A
r^{2}+C-\frac{\mu}{r}\Big)}}\,,
\end{eqnarray}
and
\begin{eqnarray}\label{a18}
u=\frac{1}{(k+1)}\sqrt{(1+\frac{\eta}{\beta}r^{2})\Big(\frac{A_{4}^{2}-\Big(\Delta+A
r^{2}+C-\frac{\mu}{r}\Big)(1+k)^{2}\Big)}{\Big(3 A
r^{2}+C+\frac{B}{1+r^{2}\gamma^{2}}\Big)}\Big)}\,,
\end{eqnarray}
respectively. For critical points we find
\begin{eqnarray}\label{a18}
r_{c}=-\frac{3}{4}\frac{\mu (k+1)^{2}}{A_{4}^{2}-(B+C)(k+1)^{2}}\,,
\end{eqnarray}
\begin{eqnarray}\label{a18}
V_{c}^{2}=\frac{\mu}{4r(B+C)-3\mu}\,,
\end{eqnarray}
and
\begin{eqnarray}\label{a18}
u_{c}=\frac{1}{4}(1+\frac{\eta}{\beta}r^{2})\frac{\Big(\frac{\mu}{r}+2Ar^{2}-\Delta+\frac{B}{1+r^{2}\gamma^{2}}\Big)}{\Big(3Ar^{2}+C+\frac{B}{1+r^{2}\gamma^{2}}\Big)}\,.
\end{eqnarray}
It is clear that $r_{c}$ depends on the equation of state parameter
$k$. This is means that the location of a critical point is not the
same for all fluids. The profiles of radial velocity, density and
accretion rate are presented in figures 8, 9 and 10 versus the
dimensionless parameter $r$ for equation of state parameter
$k=\frac{1}{2}$ and constants integration $A_{4} = 1.4$ and $A_{3} =
1$ respectively. The dotted, dashed and solid curves
 are corresponding to cases with $B=0$, $B=-0.05$ and $B=-0.1$ respectively.

The left panel of figure 8 represents the radial velocity versus the
radius. The locations of the critical points are marked by solid circles
and their coordinates are stated in the right panel. The fluid has zero
radial velocity far from the black hole and flows at the sub-sonic
speed before the critical points. In a critical point, the speed of
flow matches the speed of sound. After passing this point, in the
vicinity of the black hole the speed of flow increases and turns to
the super-sonic domain because of strong gravity. It can be observed that
velocity decreases by increasing the Horndeski/Galileon
correction factor  $B$ and the loci of the critical points get
shifted to the black hole. Therefore, the speed of infalling
particle reaches the speed of sound closer to the central mass.

The density profile of the fluid around the black hole for
different values of $B$ is shown in figure 9. Increasing the
value of the parameter $B$ increases the density. In addition, for
such isothermal fluids, the mass of the black hole changes with
time by the following relation

\begin{eqnarray}\label{a18}
\dot{M}=4\pi A_{1}(p+\rho)\sqrt{B+C}M^{2}\,.
\end{eqnarray}


\begin{figure}\label{a8}
\begin{center}
\includegraphics{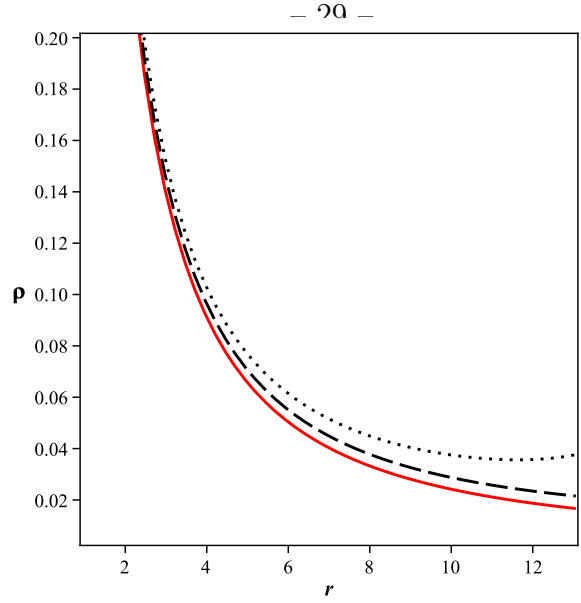}
\end{center}
\vspace{6cm} \caption{Density profile as a function of the
dimensionless parameter $r$ for the equation of state parameter
$k=\frac{1}{2}$ and constants of integration as $A_{4}=1.4$ and
$A_{3}=1$. The dotted, dashed and solid curves are corresponding to
the cases with $B=0$, $B=-0.05$ and $B=-0.1$ respectively. }
\end{figure}


\begin{figure}\label{a9}
\begin{center}
\includegraphics{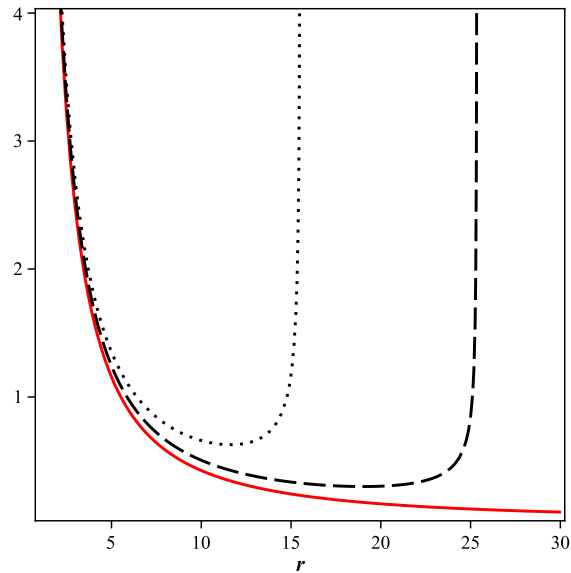}
\end{center}
\vspace{7cm} \caption{Accretion rate of isothermal fluid with
equation of state parameter $k=\frac{1}{2}$ versus the
dimensionless radial parameter $r$ for different values of the
parameter $B$. The constant parameters of integration are set
$A_{4}=1.4$ and $A_{3}=1$. The dotted, dashed and solid curves are
corresponding to the cases with $B=0$, $B=-0.05$ and $B=-0.1$ respectively. }
\end{figure}

We see that accretion rate for a general spherically symmetric static black hole
in Horndeski/Galileon gravity is different from the case of a
Schwarzschild black hole. In Horndeski/Galileon case, the accretion rate
completely depends on the nature of the accreting fluid and also the
metric parameter (here, parameter $B$). So $\dot{M}>0$ for a normal fluid which satisfies
$(p+\rho)>0$. The change of accretion rate for different values of the parameter $B$ with the same
equation of state parameter are shown in figure 10. It is seen
that accretion rate is higher in the vicinity of the black hole
because of strong gravitational effect. Also, increment of the
parameter $B$ enhances the accretion rate. By using equation (57), the
critical accretion time and the mass of the black hole are given by
\begin{eqnarray}\label{a18}
t_{cr}=[4\pi A_{1}(p+\rho)\sqrt{B+C}M_{i}]^{-1}
\end{eqnarray}
and
\begin{eqnarray}\label{a18}
M_{t}=M_{i}\Big [1-4\pi A_{1}(p+\rho)\sqrt{B+C}M_{i}t \Big ]^{-1}\,,
\end{eqnarray}
respectively. We see that for normal fluid, the
black hole mass increases by accretion matter subject to
Horndeski/Galileon gravity and increasing correction factor $B$ will
increase the black hole mass further. This behavior is studied by
Rodrigues et al. (2009) for a Schwarzschild black hole in the presence
of a non-minimally coupled scalar filed. They found that for black
hole with initial masses smaller than a certain critical value, the
accretion of the scalar filed can led to mass decreasing even in
the absence of Hawking radiation and phantom energy. Also the
black holes with initial masses greater than critical value grow by
accreting the scalar filed similar to the minimally coupled scalar case.

\newpage
\section{Summary and Conclusion}
In this paper, the geodesic motion and accretion process of a test
particle in a subclass of the general Horndeski/Galileon gravity theories in the
equatorial plane of a non-rotating black hole are investigated. In
this framework, the circular geodesics, the stability of such
orbits, oscillations under the action of small perturbations,
unstable orbits and finally accretion process of the fluid flowing
around the black hole have been investigated in a general form.
Expressions for the effective potential, energy, momentum,
characteristic radii, emission rate, epicyclic frequencies and
dynamical parameters of the system and also the mass evolution of
the black hole are derived in details. Then isothermal fluid with
equation of state $p=k \rho$ is considered and some discussions are
done for this subclass of solutions of
the Horndeski/Galileon black holes. In this manner, the metric
parameters with some approximations are obtained as $h(r)\approx
C-\frac{\mu}{r}+B$ and $f(r)\approx\frac{h(r)}{B+C}$, where by
assumption we have set $C=1$. The effect of the Horndeski/Galileon
correction factor $B$ is considered for each case and our solutions
are compared with the Schwarzschild black hole solutions. Our
analysis has revealed that these Horndeski/Galileon solutions have
deviations from the Schwarzschild solutions (which is recovered
where $B=0$) substantially. Our results show that Horndeski/Galileon
correction factor affects the effective potential and as a result
changes the loci of the stable and unstable circular orbits. For
larger values of this parameter, $V_{eff}$ achieves larger value and
unstable circular orbits will be located at smaller radii, whereas
stable orbits will be located at farther distances from the central
mass. When the metric parameter $B$ decreases, two points joint
together at ISCO.

In this spacetime the location of the characteristic radii such as
$r_{isco}$, $r_{ph}$, $r_{sing}$ and $r_{mb}$ have considerable
deviation from the Schwarzschild solutions. These radii, except
$r_{mb}$, are decreasing functions with respect to the
Horndeski/Galileon correction factor $B$, that is, they will be
closer to the central mass for larger values of $B$. As $r_{isco}$
represents the inner edge of the accretion disk, our results show
that for larger deviations, the disk will be extended close to the
central mass. The behavior of $r_{mb}$ is different and has a minima
at the Schwarzschild case. This radii is a decreasing function for
$B\subset(-1, 0)$  and it is a growing function for $B\subset (0,
1.125)$, and finally it coincides with the innermost stable circular
orbit.

As Horndeski/Galileon correction factor $B$ grows, the energy raises whereas angular momentum
decreases and one can see from the energy diagram that the range of
bound orbit will be smaller for larger deviations. Increasing the
parameter $B$ enhances the $E_{isco}$ and then the efficiency of
accretion will be decreased accordingly. In the case of Schwarzschild black hole
the efficiency equals to $0.057$ at $r = 6r_{g}$. For negative values
of the correction factor $B$, efficiency grows up and becomes greater in the
Schwarzschild black hole limit ($B=0$), where this behavior is reverse for
positive values of the correction factor. The flux of the radiation energy
has a maximum in the vicinity of the black hole and decreases
at smaller radii. By increasing the parameter $B$, the flux raises and
the maximum of the flux turns to the smaller radii but the reverse
happens after this point. The dependence on this parameter is very
considerable in the vicinity of the black hole, but it is weak far
from the black hole. These behaviors are the same for temperature.

In this paper, in addition to investigation of the circular orbits and
their properties, epicyclic frequencies are studied.
Vertical epicyclic frequency is monotonically decreasing function
of $r$ and has no extrema. Deviation from the Schwarzschild case
in this Horndeski/Galileon setup has no effect on the vertical epicyclic frequency, while the radial
epicyclic frequency always has a maximum and the effect of
Horndeski/Galileon correction factor on it is considerable.
Dependence on this parameter is significant close to the black hole,
but far from the central mass this effect is weak. Increasing the
parameter $B$ shifts the maximum to the smaller radii. We found that
$\Omega_{r}<\Omega_{\theta}$ and the ratio of
$\frac{\Omega_{\theta}}{\Omega_{r}}$ is decreasing function with
respect to $r$. In the vicinity of the black hole, this ratio is
very larger than unity but far from the black hole, it turns to the unity
and it decreases by increasing $B$. The dependence of some important
resonances such as parametric and forced resonances with respect to
Horndeski/Galileon parameter $B$ is obtained, which this dependence is a monotonically
decreasing function, that is, larger values of the Horndeski/Galileon
correction factor lead to happening the resonance at the smaller
radii.

Finally, the accretion process of the isothermal fluid is discussed
for equation of state parameter $k=\frac{1}{2}$ and the behavior of the radial velocity and density
are studied. Radial velocity is a decreasing function of $r$ as fluids have zero radial velocity in far from the black hole.
When accretion happens, fluid passes at a critical point where in this
point, the speed of flow matches the speed of sound. The fluid flows
at sub-sonic speed before the critical point. After passing this
point and in the vicinity of the black hole, because of strong gravitational field, the
speed of flow increases and will be super-sonic then after.
Our results show that velocity decreases by increasing $B$ and the loci of the critical
point shifts towards the black hole. Therefore, the speed of infalling
particles reaches the speed of sound closer to the central mass.
Finally, the rate of accretion is discussed where we found that
this rate depends on the nature of the fluid and also the
metric parameter. For normal fluid, $\dot{M}>0$ and its value is
larger in the vicinity of the black hole because of strong
gravitational effect and positive deviation from the Schwarzschild case
increases the accretion rate.

We note that in this paper a non-spinning particle is considered.  When a particle has spin, this spin
has considerable influence on the particle's orbit.  Also, a perfect fluid is considered
and viscosity as well as magnetic field of the accretion disk are ignored
for simplicity. However, these effects can affect the motion of the test
particle and therefore they can affect the structure and the
emission rate of the accretion disk. So, we are going to study the
behavior of spinning particles and accretion of viscose fluids
subject to the Horndeski/Galileon gravity in presence of a magnetic field
in our future work.\\

{\bf Acknowledgement}\\
We would like to thank the referee for insightful comments. The work of K. Nozari has been financially supported by Research Institute for
Astronomy and Astrophysics of Maragha (RIAAM) under research project No. 1/5411-6.


\end{document}